\documentclass[aps,prc,twocolumn,showpacs,superscriptaddress]{revtex4-2} 

\usepackage{graphicx}  
\usepackage{dcolumn}   
\usepackage{bm}        
\usepackage{amssymb}   
\usepackage{xcolor}    
\usepackage{lipsum}
\usepackage{wrapfig}

\begin{document}

\title{Multidimensional Measurements of Beam Single Spin Asymmetries in Semi-inclusive Deep-inelastic Charged Kaon Electroproduction off Protons in the Valence Region}


\newcommand*{\ANL}{Argonne National Laboratory, Argonne, Illinois 60439}
\newcommand*{\ANLindex}{1}
\affiliation{\ANL}
\newcommand*{\ASU}{Arizona State University, Tempe, Arizona 85287-1504}
\newcommand*{\ASUindex}{2}
\affiliation{\ASU}
\newcommand*{\CSUDH}{California State University, Dominguez Hills, Carson, CA 90747}
\newcommand*{\CSUDHindex}{3}
\affiliation{\CSUDH}
\newcommand*{\CANISIUS}{Canisius University, Buffalo, NY}
\newcommand*{\CANISIUSindex}{4}
\affiliation{\CANISIUS}
\newcommand*{\SACLAY}{IRFU, CEA, Universit'{e} Paris-Saclay, F-91191 Gif-sur-Yvette, France}
\newcommand*{\SACLAYindex}{5}
\affiliation{\SACLAY}
\newcommand*{\CNU}{Christopher Newport University, Newport News, Virginia 23606}
\newcommand*{\CNUindex}{6}
\affiliation{\CNU}
\newcommand*{\UCONN}{University of Connecticut, Storrs, Connecticut 06269}
\newcommand*{\UCONNindex}{7}
\affiliation{\UCONN}
\newcommand*{\DUKE}{Duke University, Durham, North Carolina 27708-0305}
\newcommand*{\DUKEindex}{8}
\affiliation{\DUKE}
\newcommand*{\DUQUESNE}{Duquesne University, 600 Forbes Avenue, Pittsburgh, PA 15282 }
\newcommand*{\DUQUESNEindex}{9}
\affiliation{\DUQUESNE}
\newcommand*{\FU}{Fairfield University, Fairfield CT 06824}
\newcommand*{\FUindex}{10}
\affiliation{\FU}
\newcommand*{\FERRARAU}{Universita' di Ferrara , 44121 Ferrara, Italy}
\newcommand*{\FERRARAUindex}{11}
\affiliation{\FERRARAU}
\newcommand*{\FIU}{Florida International University, Miami, Florida 33199}
\newcommand*{\FIUindex}{12}
\affiliation{\FIU}
\newcommand*{\FSU}{Florida State University, Tallahassee, Florida 32306}
\newcommand*{\FSUindex}{13}
\affiliation{\FSU}
\newcommand*{\GWUI}{The George Washington University, Washington, DC 20052}
\newcommand*{\GWUIindex}{14}
\affiliation{\GWUI}
\newcommand*{\GSIFFN}{GSI Helmholtzzentrum fur Schwerionenforschung GmbH, D-64291 Darmstadt, Germany}
\newcommand*{\GSIFFNindex}{15}
\affiliation{\GSIFFN}
\newcommand*{\INFNFE}{INFN, Sezione di Ferrara, 44100 Ferrara, Italy}
\newcommand*{\INFNFEindex}{16}
\affiliation{\INFNFE}
\newcommand*{\INFNFR}{INFN, Laboratori Nazionali di Frascati, 00044 Frascati, Italy}
\newcommand*{\INFNFRindex}{17}
\affiliation{\INFNFR}
\newcommand*{\INFNGE}{INFN, Sezione di Genova, 16146 Genova, Italy}
\newcommand*{\INFNGEindex}{18}
\affiliation{\INFNGE}
\newcommand*{\INFNRO}{INFN, Sezione di Roma Tor Vergata, 00133 Rome, Italy}
\newcommand*{\INFNROindex}{19}
\affiliation{\INFNRO}
\newcommand*{\INFNTUR}{INFN, Sezione di Torino, 10125 Torino, Italy}
\newcommand*{\INFNTURindex}{20}
\affiliation{\INFNTUR}
\newcommand*{\INFNCAT}{INFN, Sezione di Catania, 95123 Catania, Italy}
\newcommand*{\INFNCATindex}{21}
\affiliation{\INFNCAT}
\newcommand*{\INFNPAV}{INFN, Sezione di Pavia, 27100 Pavia, Italy}
\newcommand*{\INFNPAVindex}{22}
\affiliation{\INFNPAV}
\newcommand*{\ORSAY}{Universit'{e} Paris-Saclay, CNRS/IN2P3, IJCLab, 91405 Orsay, France}
\newcommand*{\ORSAYindex}{23}
\affiliation{\ORSAY}
\newcommand*{\JMU}{James Madison University, Harrisonburg, Virginia 22807}
\newcommand*{\JMUindex}{24}
\affiliation{\JMU}
\newcommand*{\KSU}{King Saud University, Riyadh, Kingdom of Saudi Arabia}
\newcommand*{\KSUindex}{25}
\affiliation{\KSU}
\newcommand*{\KNU}{Kyungpook National University, Daegu 41566, Republic of Korea}
\newcommand*{\KNUindex}{26}
\affiliation{\KNU}
\newcommand*{\ULS}{Universidad de La Serena, Avda. Juan Cisternas 1200, La Serena, Chile}
\newcommand*{\ULSindex}{27}
\affiliation{\ULS}
\newcommand*{\LAMAR}{Lamar University, 4400 MLK Blvd, PO Box 10046, Beaumont, Texas 77710}
\newcommand*{\LAMARindex}{28}
\affiliation{\LAMAR}
\newcommand*{\MIT}{Massachusetts Institute of Technology, Cambridge, Massachusetts  02139-4307}
\newcommand*{\MITindex}{29}
\affiliation{\MIT}
\newcommand*{\MISS}{Mississippi State University, Mississippi State, MS 39762-5167}
\newcommand*{\MISSindex}{30}
\affiliation{\MISS}
\newcommand*{\NUNA}{School of Physics and Institute for Nonperturbative Physics, Nanjing University, Nanjing 210093, Jiangsu, China}
\newcommand*{\NUNAindex}{31}
\affiliation{\NUNA}
\newcommand*{\NUPTNA}{School of Science, Nanjing University of Posts and Telecommunications, Nanjing 210023, Jiangsu, China}
\newcommand*{\NUPTNAindex}{32}
\affiliation{\NUPTNA}
\newcommand*{\UNH}{University of New Hampshire, Durham, New Hampshire 03824-3568}
\newcommand*{\UNHindex}{33}
\affiliation{\UNH}
\newcommand*{\OHIOU}{Ohio University, Athens, Ohio  45701}
\newcommand*{\OHIOUindex}{34}
\affiliation{\OHIOU}
\newcommand*{\ODU}{Old Dominion University, Norfolk, Virginia 23529}
\newcommand*{\ODUindex}{35}
\affiliation{\ODU}
\newcommand*{\JLUGiessen}{II Physikalisches Institut der Universitaet Giessen, 35392 Giessen, Germany}
\newcommand*{\JLUGiessenindex}{36}
\affiliation{\JLUGiessen}
\newcommand*{\ROMAII}{Universita' di Roma Tor Vergata, 00133 Rome Italy}
\newcommand*{\ROMAIIindex}{37}
\affiliation{\ROMAII}
\newcommand*{\SDU}{Shandong University, Qingdao, Shandong 266237, China}
\newcommand*{\SDUindex}{38}
\affiliation{\SDU}
\newcommand*{\MSU}{Skobeltsyn Institute of Nuclear Physics, Lomonosov Moscow State University, 119234 Moscow, Russia}
\newcommand*{\MSUindex}{39}
\affiliation{\MSU}
\newcommand*{\SCAROLINA}{University of South Carolina, Columbia, South Carolina 29208}
\newcommand*{\SCAROLINAindex}{40}
\affiliation{\SCAROLINA}
\newcommand*{\SUNA}{School of Physics, Southeast University, Nanjing 211189, Jiangsu, China}
\newcommand*{\SUNAindex}{41}
\affiliation{\SUNA}
\newcommand*{\TEMPLE}{Temple University,  Philadelphia, PA 19122 }
\newcommand*{\TEMPLEindex}{42}
\affiliation{\TEMPLE}
\newcommand*{\JLAB}{Thomas Jefferson National Accelerator Facility, Newport News, Virginia 23606}
\newcommand*{\JLABindex}{43}
\affiliation{\JLAB}
\newcommand*{\UTFSM}{Universidad T\'{e}cnica Federico Santa Mar\'{i}a, Casilla 110-V Valpara\'{i}so, Chile}
\newcommand*{\UTFSMindex}{44}
\affiliation{\UTFSM}
\newcommand*{\INSUBRIA}{Universit\`{a} degli Studi dell'Insubria, 22100 Como, Italy}
\newcommand*{\INSUBRIAindex}{45}
\affiliation{\INSUBRIA}
\newcommand*{\BRESCIA}{Universit`{a} degli Studi di Brescia, 25123 Brescia, Italy}
\newcommand*{\BRESCIAindex}{46}
\affiliation{\BRESCIA}
\newcommand*{\UCR}{University of California Riverside, 900 University Avenue, Riverside, CA 92521, USA}
\newcommand*{\UCRindex}{47}
\affiliation{\UCR}
\newcommand*{\URICH}{University of Richmond, 138 UR Drive, Richmond, VA 23173, USA}
\newcommand*{\URICHindex}{48}
\affiliation{\URICH}
\newcommand*{\GLASGOW}{University of Glasgow, Glasgow G12 8QQ, United Kingdom}
\newcommand*{\GLASGOWindex}{49}
\affiliation{\GLASGOW}
\newcommand*{\UTK}{University of Tennessee, Knoxville, Tennessee 37996, USA}
\newcommand*{\UTKindex}{50}
\affiliation{\UTK}
\newcommand*{\YORK}{University of York, York YO10 5DD, United Kingdom}
\newcommand*{\YORKindex}{51}
\affiliation{\YORK}
\newcommand*{\VIRGINIA}{University of Virginia, Charlottesville, Virginia 22901}
\newcommand*{\VIRGINIAindex}{52}
\affiliation{\VIRGINIA}
\newcommand*{\WM}{College of William and Mary, Williamsburg, Virginia 23187-8795}
\newcommand*{\WMindex}{53}
\affiliation{\WM}
\newcommand*{\YEREVAN}{Yerevan Physics Institute, 375036 Yerevan, Armenia}
\newcommand*{\YEREVANindex}{54}
\affiliation{\YEREVAN} 

\newcommand*{\NOWJLAB}{Thomas Jefferson National Accelerator Facility, Newport News, Virginia 23606}


\author {A.~Kripko} 
\affiliation{\JLUGiessen}
\author {S.~Diehl} 
\affiliation{\JLUGiessen}
\affiliation{\UCONN}
\author {K.~Joo} 
\affiliation{\UCONN}

\author {P.~Achenbach} 
\affiliation{\JLAB}
\author {J. S. Alvarado} 
\affiliation{\ORSAY}
\author {M.~Amaryan}
\affiliation{\ODU}
\author {W.R.~Armstrong} 
\affiliation{\ANL}
\author {H.~Atac} 
\affiliation{\TEMPLE}
\author {H.~Avakian} 
\affiliation{\JLAB}
\author {L.~Baashen} 
\affiliation{\KSU}
% ...

\collaboration{The CLAS Collaboration}
\noaffiliation

\begin{abstract}
Measurements of beam single spin asymmetries in semi-inclusive deep inelastic electron scattering (SIDIS) with positively charged kaons off protons have been performed with 10.6 and 10.2 GeV incident electron beams using the CLAS12 spectrometer at Jefferson Lab. We report an analysis of the electroproduction of positively charged kaons over a large kinematic range of fractional energy, Bjorken $x$, transverse momentum, and photon virtualities $Q^2$ ranging from 1 GeV$^2$ up to 6 GeV$^2$. This is the first published multi-dimensionally binned CLAS12 measurement of a kaon SIDIS single spin asymmetry in the valence quark regime. The data provide constraints on the structure function ratio $F_{LU}^{\sin{\phi}}/F_{UU}$, where $F_{LU}^{\sin{\phi}}$ is a quantity with a leading twist of twist-3 that can reveal novel aspects of the quark-gluon correlations within the nucleon. The impact of the data on understanding the underlying reaction mechanisms and their kinematic variation is explored using theoretical models for the different contributing twist-3 parton distribution functions (PDFs) and fragmentation functions (FFs).

\end{abstract}


\maketitle

\section{Introduction}

Most of the visible matter in the Universe consists of nucleons. Their mass and spin emerge from the strong interactions of their constituents, meaning that the description of their internal dynamics is important to understand the theory of strong interactions.

To better understand the nucleon's structure, a focus of the hadron physics community has moved beyond the collinear parton distribution functions (PDFs) \cite{GAO20181,doi:10.1146/annurev-nucl-011720-042725,RevModPhys.92.045003} towards the study of the partons' motion and their spatial distribution in the transverse plane, perpendicular to the momentum of the parent hadron, via deep inelastic scattering (DIS) of lepton beams off nucleons. PDFs encode information about the momentum-dependent distribution of quarks inside the nucleon at twist-2. The contribution of twist-t operators in the process amplitude is suppressed by 2-$t$ of the hard scale of the process through which the structure of the nucleon is studied \cite{jaffe1996spintwisthadronstructure}. Higher-twist functions do not have a parton model interpretation, as they are correlations between quarks and gluons that describe the non-local parts of QCD \cite{ELLIS198329}.

In semi-inclusive DIS (SIDIS), at least one specified hadron in the final state besides the scattered lepton is measured. SIDIS is a well-established tool for studying nucleon structural distributions in the plane transverse to its light-front (longitudinal) direction of motion and for studying collinear PDFs. Spin asymmetries in polarized SIDIS are related to convolutions of PDFs with fragmentation functions (FFs), and thus are of great interest among the hadron physics community \cite{PhysRevD.43.261,PhysRevD.67.114014,PhysRevD.70.117504,Metz2004,PhysRevD.87.094019,Avakian_2019,ANSELMINO2020103806}. The FFs can be interpreted in twist-2 as the probability for a given struck parton to emit a particular hadron \cite{METZ2016136}. 

Sizable single spin asymmetries (SSAs) for charged hadrons have been measured by COMPASS \cite{ADOLPH20141046}, for charged pions by CLAS \cite{PhysRevD.89.072011} and CLAS12 \cite{PhysRevLett.128.062005} using multidimensional binning. The HERMES \cite{PhysRevLett.84.4047,PhysRevD.64.097101,AIRAPETIAN2007164,2019134886} experiment also measured SSAs for hadrons, including kaons, in a similar kinematic range, but with less statistics. The beam SSAs are twist-3 objects, which means that they are suppressed by $\mathcal{O}(M/Q)$, where $M$ is the mass of the target nucleon and $Q^2$ is the virtuality of the photon.

\begin{figure}[h!]
\begin{center}
    \includegraphics[width=0.49\textwidth]{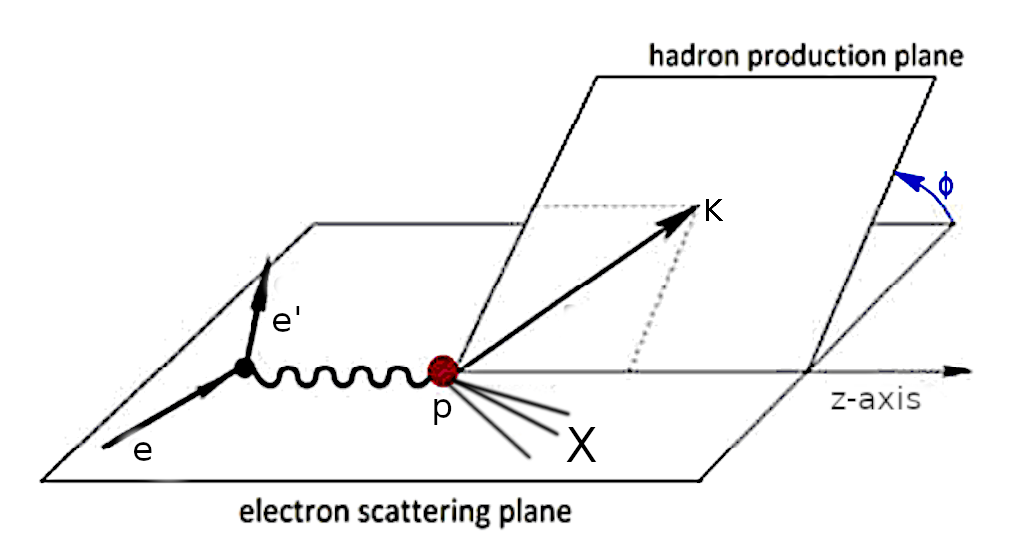}
	\caption{Schematic illustration of the reaction kinematics of the single kaon semi-inclusive deep inelastic scattering process.}
	\label{fig:SIDIS_kin}
	\end{center}
\end{figure}

In this Letter we present high statistics beam SSAs measured in $K^{+}$ SIDIS of longitudinally polarized electrons off unpolarized protons with a wide range of multidimensional kinematics on $Q^2\in[1,6]$ GeV$^2$, $x_B\in[0.06,0.6]$, $z\in[0.1,0.7]$ and $P_T\in[0.1,0.8]$ GeV. Here $Q^2$ is the momentum transferred into the system by the lepton probe (the photon virtuality), $x_B$ is the fraction of the proton’s momentum carried by the struck quark at leading twist, $P_T$ is the transverse momentum of the hadron with respect to the virtual photon, and $z$ is the fraction of the virtual photon’s energy carried by the outgoing hadron in the lab frame. A schematic illustration of the process is shown in Fig. \ref{fig:SIDIS_kin}. A definition of these relevant kinematic variables ($Q^2$, $x_B$, $z$, $P_T$) can be found in Ref. \cite{DIEHL2023104069}. Due to their long lifetime and the resulting possibility for direct detection, pseudo-scalar mesons such as $\pi^+$ and $K^+$ are especially suited for SIDIS measurements.

In the one-photon exchange approximation, the beam SSAs are defined as \cite{DIEHL2023104069}:

\begin{eqnarray}
    A_{LU}(z,P_T, x_B,Q^2,\phi)=\frac{d\sigma^+-d\sigma^-}{d\sigma^++d\sigma^-}=\nonumber\\
    \frac{A_{LU}^{\sin{\phi}}\sin{\phi}}{1+A_{UU}^{\cos{\phi}}\cos{\phi}+A_{UU}^{\cos{2\phi}}\cos{2\phi}},
\label{eq:ALU}
\end{eqnarray}
where $d\sigma^{\pm}$ are the differential cross sections for the respective helicity state, {\it i.e.} the spin of the electron is parallel or antiparallel to the beam direction. The $L$ and $U$ subscripts represent the longitudinally polarized and the unpolarized states of the beam and the target, respectively. $\phi$ is the azimuthal angle between the electron scattering plane and the hadronic reaction plane as shown in Fig.~\ref{fig:SIDIS_kin}.

In the following, we will focus on the $\sin{\phi}$ moment, $A_{LU}^{\sin{\phi}}$, which provides access to dynamical aspects of proton structure, as it is proportional to the polarized structure function ratio $F_{LU}^{\sin{\phi}}/F_{UU}$ \cite{DIEHL2023104069}:

\begin{equation}
    \frac{F_{LU}^{\sin{\phi}}}{F_{UU}}=\frac{F_{LU}^{\sin{\phi}}}{F_{UU,T}+\epsilon F_{UU,L}}=\frac{A_{LU}^{\sin{\phi}}}{\sqrt{2\epsilon(1-\epsilon)}}.
\label{eq:FLU}
\end{equation}

\noindent Here, $F_{UU,T}$ and $F_{UU,L}$ are the contributions from the longitudinal and transverse polarizations of the virtual photon, respectively, with $\epsilon$ being the ratio of their fluxes.

Although factorization has not yet been proven in connection with twist-three observables \cite{BACCHETTA2019134850}, if one assumes that it is valid, then our data could be interpreted in terms of TMDs. The TMDs and FFs involved for given quark flavors can be obtained from the measurement of different mesons in the SIDIS final state (flavor decomposition). Moreover, the TMDs for distinct quark flavors may show different kinematic dependencies, making a flavor separation in fully differential kinematics essential. In this context kaon SIDIS plays an essential role in identifying the behavior of strange quark TMDs and FFs and their role in the different mechanisms.
\noindent
If one stays in the kinematic regime where the factorized convolution formula is assumed to be valid (small $P_T$) and one can write \cite{AlessandroBacchetta,LEVELT1994357}:

\begin{eqnarray}
F_{LU}^{\sin{\phi}}=\frac{2M}{Q}\kappa\Biggl[-\frac{\boldsymbol{\hat{h}k_T}}{M_h}\left(xeH_1^{\perp}+\frac{M_h}{M}f_1\frac{\tilde{G^{\perp}}}{z}\right)\nonumber\\
+\frac{\boldsymbol{\hat{h}p_T}}{M}\left(xg^{\perp}D_1+\frac{M_h}{M}h_1^{\perp}\frac{\tilde{E}}{z}\right)\Biggr].
\label{eq:FLUsin}
\end{eqnarray}

\noindent Here $\kappa$ denotes the convolution of PDFs and FFs weighted by a kinematic factor, $e$ is a twist-3 PDF, $H_1^{\perp}$ is called the Collins FF, $f_1$ is the unpolarized distribution function, $\tilde{G^{\perp}}$ is a twist-3 FF, $g^{\perp}$ is a twist-3 T-odd distribution function, $D_1$ is the unpolarized FF, $h_1^{\perp}$ is the Boer-Mulders function \cite{PhysRevD.57.5780} and $\tilde{E}$ is a twist-3 FF. More details of these functions can be found in Refs. \cite{METZ2016136,ANSELMINO2020103806,AIRAPETIAN2007164}. $\boldsymbol{p_T}$ and $\boldsymbol{k_T}$ are the intrinsic quark transverse momentum in the distribution and fragmentation functions, respectively. $\boldsymbol{\hat{h}}$ is a unit vector in the direction of the kaon's transverse momentum and $M_h$ is the kaon mass. Most twist-3 structure functions can be separated into 3 terms using the equations of motion derived from the underlying QCD theory: a twist-2 term, related to a single-parton density, a twist-3 part, which contains information on quark-gluon correlations \cite{PhysRevD.85.014021,Schweitzer2013} and a term proportional to the current-quark mass, which is usually neglected for light quarks. The structure function $F_{LU}^{\sin{\phi}}$ contains only terms where either the PDF or the FF is twist-3, and is therefore sensitive to quark-gluon correlations \cite{inproceedings}. As perturbative QCD could not describe the several percent magnitude of the experimentally observed asymmetry (see Ref. \cite{Chiappetta1986}), non-perturbative mechanisms have been proposed. The first one involves the $eH_1^{\perp}$ term \cite{Anatoli_Vasilievich_Efremov_2003,cebulla2007twist3}, attributing the asymmetry to a coupling between the Collins FF $H_1^{\perp}$ and $e$ \cite{Accardi2020,PhysRevD.85.014021,Schweitzer2013}. Although the collinear PDF $e(x)$ does not have a direct probabilistic interpretation due to its twist-3 nature, its moments provide insights to the contribution to the nucleon mass from the finite quark masses and to the transverse force experienced by a transversely polarized quark in an unpolarized nucleon immediately after scattering \cite{2019134886}. The second mechanism involves convolution of the Boer-Mulders function $h_1^{\perp}$ with the FF $\tilde{E}$ and the coupling between the unpolarized distribution function $f_1$ and the twist-3 FF $\tilde{G^{\perp}}$. In addition to those mentioned above, a mechanism involving the poorly known twist-3 TMD $g^{\perp}$ can also generate the beam SSA \cite{DIEHL2023104069}. $g^{\perp}$ is sensitive to target quark-gluon correlations and it is related to the Boer-Mulders PDF through the QCD equations of motion \cite{2019134886}. During the subsequent discussion, the main focus will be on the $eH_1^{\perp}$ and $g^{\perp}D_1$ terms in Eq. \ref{eq:FLUsin}. An overview of the previous measurements of these TMDs and FFs can be found in Ref. \cite{Avakian2016}.

\section{Experimental setup}

SIDIS positively charged kaon electroproduction was measured at Jefferson Lab with CLAS12 (CEBAF Large Acceptance Spectrometer for experiments at 12 GeV) \cite{BURKERT2020163419}. Beam SSAs were extracted over a wide range in $Q^2$, $x_B$, $z$, $P_T$ and $\phi$. The incident electron beam was longitudinally polarized and the target was unpolarized liquid hydrogen. Data were taken with two different beam energies of 10.6 and 10.2~GeV. The CLAS12 forward detector consists of six identical sectors within a toroidal magnetic field. The momentum and the charge of the particles were determined by three regions of drift chambers from the curvature of the particle trajectories in the magnetic field. The electron identification was based on a lead-scintillator electromagnetic sampling calorimeter in combination with a Cherenkov counter. The identification of charged kaons is based on time-of-flight measurements. Although the dedicated time-of-flight system is able to achieve a 4$\sigma$ kaon-pion separation up to 2.8 GeV momentum, there is still a non-negligible pion contamination in the kaon sample due to the high pion-to-kaon ratio. A deep neural network was developed, which combines time information and the deposited energies obtained from multiple detector components and takes their correlations into account. The neural network consists of 3 fully connected hidden layers with 128 neurons per layer. This way the kaons could be identified more reliably up to 3 GeV momentum as the average purity was increased to 80-90\% from 60-70\% according to Monte Carlo simulations, while keeping more than 50\% of the total sample. These results could only be achieved for positive kaons, as the negative ones have smaller statistics and higher pion contamination. For this reason the negative kaons are not considered in this letter. For the selection of deeply inelastic scattered electrons, constraints on $Q^2 > 1$ GeV$^2$, on the energy fraction of the incoming lepton carried by the virtual photon $y < 0.75$, on Feynman x $x_F>0$, and on the invariant mass of the hadronic final state $W > 2$ GeV were applied. Furthermore, it was required that the $e'K^{+}X$ missing mass be larger than 1.6 GeV to reduce the contribution from exclusive channels. The only exclusive channel that is above this cut and visible in the missing mass spectrum is the $\Lambda(1890)$, but it is responsible for less than 1\% of the sample.  

\begin{figure}[h!]
\begin{center}
    \includegraphics[width=0.495\textwidth]{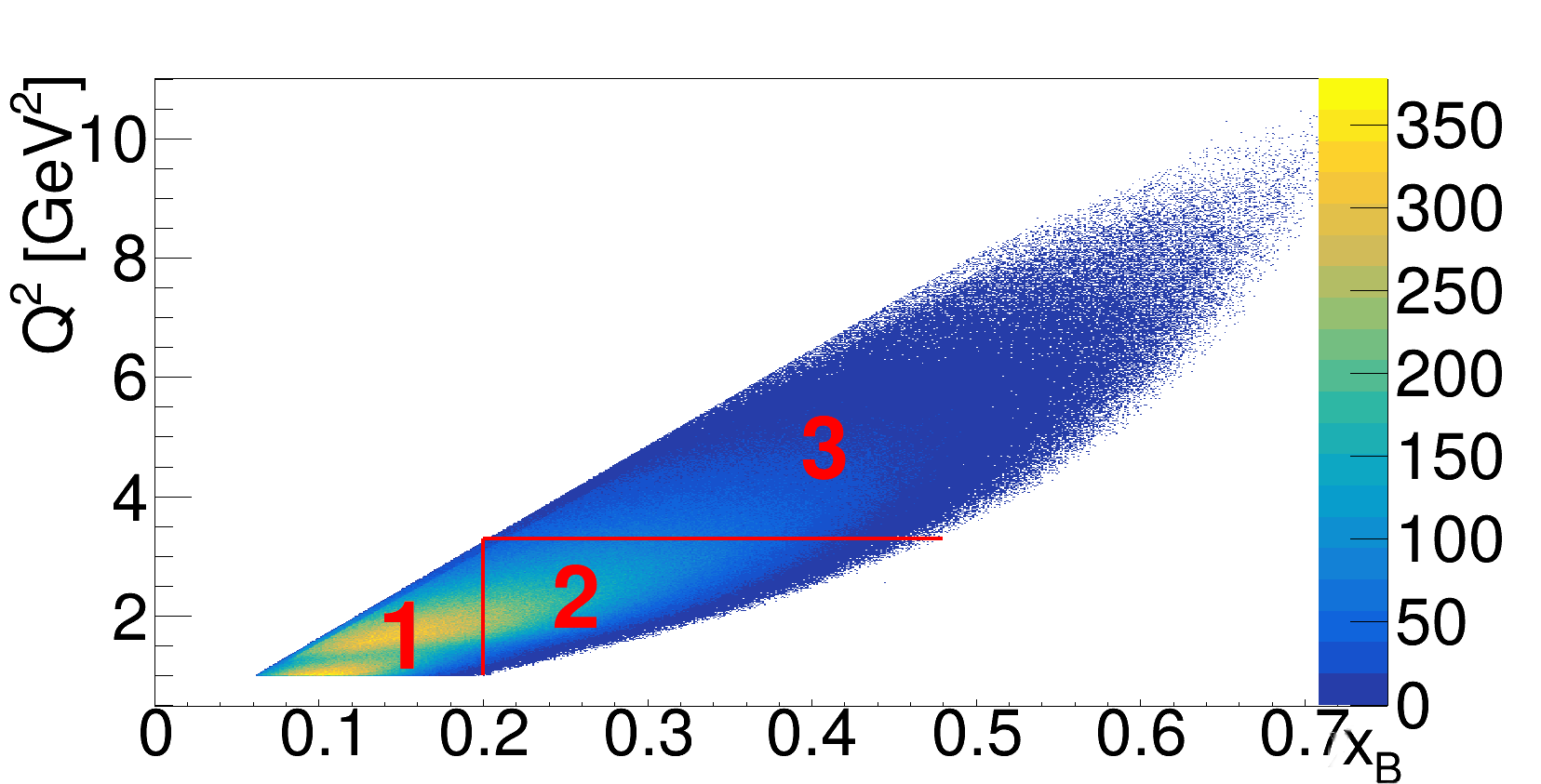}
	\caption{Distribution of $Q^2$ versus $x_B$ with the bin boundaries.}
	\label{fig:Q2_x_binning}
	\end{center}
\end{figure}

Although the model-independent structure function ratio $F_{LU}^{\sin{\phi}}/F_{UU}$ was studied at HERMES \cite{Airapetian_2019}, CLAS \cite{PhysRevD.89.072011} and CLAS12 \cite{PhysRevLett.128.062005} for different mesons in the final state, over the last 20 years, there is still no consistent understanding of the contribution of each part of Eq. \ref{eq:FLUsin} to the total structure function. The high statistics on an extended kinematic range, available with the new CLAS12 data, enables a precise multidimensional analysis for the first time, providing an excellent basis for the extraction of asymmetries for positively charged kaons in the valence quark regime. The data presented in this paper can be used as an additional input for global fits. For the multidimensional binning, first the electron variables are sorted in three bins on the $Q^2-x_B$-plane. This is illustrated in Fig. \ref{fig:Q2_x_binning}. In each $Q^2-x_B$-bin three bins are created in $P_T$ and $z$, and a fine binning is applied for the remaining variable. The exact bin borders can be found in the Supplemental Material \cite{supl}. In this way the dependence of $F_{LU}^{\sin{\phi}}/F_{UU}(z,P_T,x_B,Q^2)$ (Eq. \ref{eq:ALU} and \ref{eq:FLU}) on each individual kinematic variable can be examined.

The beam SSA and its statistical uncertainty were experimentally determined from the number of counts with positive and negative helicity ($N_i^{\pm}$) in a specific bin $i$ as

\begin{equation}
    A_{LU}=\frac{1}{P_e}\frac{N_i^+-N_i^-}{N_i^++N_i^-}; \sigma_{A_{LU}}=\frac{2}{P_e}\sqrt{\frac{N_i^+N_i^-}{(N_i^++N_i^-)^3}}.
\end{equation}

\noindent Here $P_e$ is the average magnitude of the beam polarization, which was measured with a M\o{}ller polarimeter upstream of CLAS12 to be 87.24\%$\pm$ 2.42\%. The polarization was flipped at 30 Hz to minimize systematic effects. The beam SSA was measured as a function of the azimuthal angle $\phi$ and was fitted with a $\sin{\phi}$ function to extract $A_{LU}^{\sin{\phi}}$.

\section{Systematic uncertainties}

Several sources of systematic uncertainty were investigated, including beam polarization ($\approx$3\%), radiative effects and the effect of the $\Lambda(1890)$ exclusive channel ($\approx$4\%), and contamination from baryon resonances ($\approx$1\%). The influence of additional $\cos{\phi}$ and $\cos{2\phi}$ azimuthal modulations on the extracted $\sin{\phi}$ amplitude was also evaluated and found to be small ($\approx$4\%). A detailed GEANT4 \cite{GEANT4:2002zbu} based Monte Carlo simulation \cite{UNGARO2020163422} was performed to study acceptance and bin-migration effects ($\approx$1\%), which were both found to be negligible compared to the other contributions. The same simulation was also used to estimate the pion contamination in the kaon sample. On the basis of this information, the pion asymmetries were subtracted from the kaon asymmetries in every kinematic bin. The differences after and before the correction were below 5\% in most kinematic bins. The estimated contamination as well as the measured pion asymmetry and the uncorrected kaon asymmetry is listed in the Supplemental Material \cite{supl}. The same kinematic binning and cuts were used to extract these pion asymmetries as for the kaons. Every other part of the analysis procedure was identical to what is described in the pion paper \cite{PhysRevLett.128.062005}. The uncertainty of the contamination determined from the simulation was estimated by repeating the extraction procedure using two different PID (Particle Identification) methods: with and without using the neural network. This source of uncertainty was found to be the highest one, slightly below 15\% in average. Another PYTHIA MC generated dataset, optimized for CLAS12, was used to estimate the contamination coming from the target fragmentation region and it was found to be negligible compared to the other uncertainty sources above 0.3 in $z$. Most kinematic bins are above this value with the exception of the lowest $z$ bins in case of the fine binning in $z$ and the lowest $z$ bin in case of the fine binning in $P_T$. Although the separation of target fragmentation in these bins is model-dependent, they are included for completeness. These bins do not affect the other ones due to the multidimensional binning. The total point-to-point systematic uncertainty of $F_{LU}^{\sin{\phi}}/F_{UU}$, defined as the square root of the quadratic sum of the uncertainties from all sources, is typically on the order of 22\%, slightly below the average statistical uncertainty, which is around 29\%. The ratio $F_{LU}^{\sin{\phi}}/F_{UU}$ was extracted and is listed in Supplemental Material \cite{supl} and in the CLAS physics database \cite{CLAS_database}, alongside the mean value of the kinematic variables in each bin. Also the average of the systematic uncertainty in the kinematic bins coming from different sources is shown in the Supplemental Material \cite{supl}. Furthermore, a comparison with the previous kaon SIDIS $F_{LU}^{\sin{\phi}}/F_{UU}$ values from HERMES and with the pion values, used to correct for pion contamination, can be found in the Supplemental Material \cite{supl}.

\section{Results and discussion}

\begin{figure*}[h!]
\begin{center}
    \includegraphics[width=0.65\textwidth]{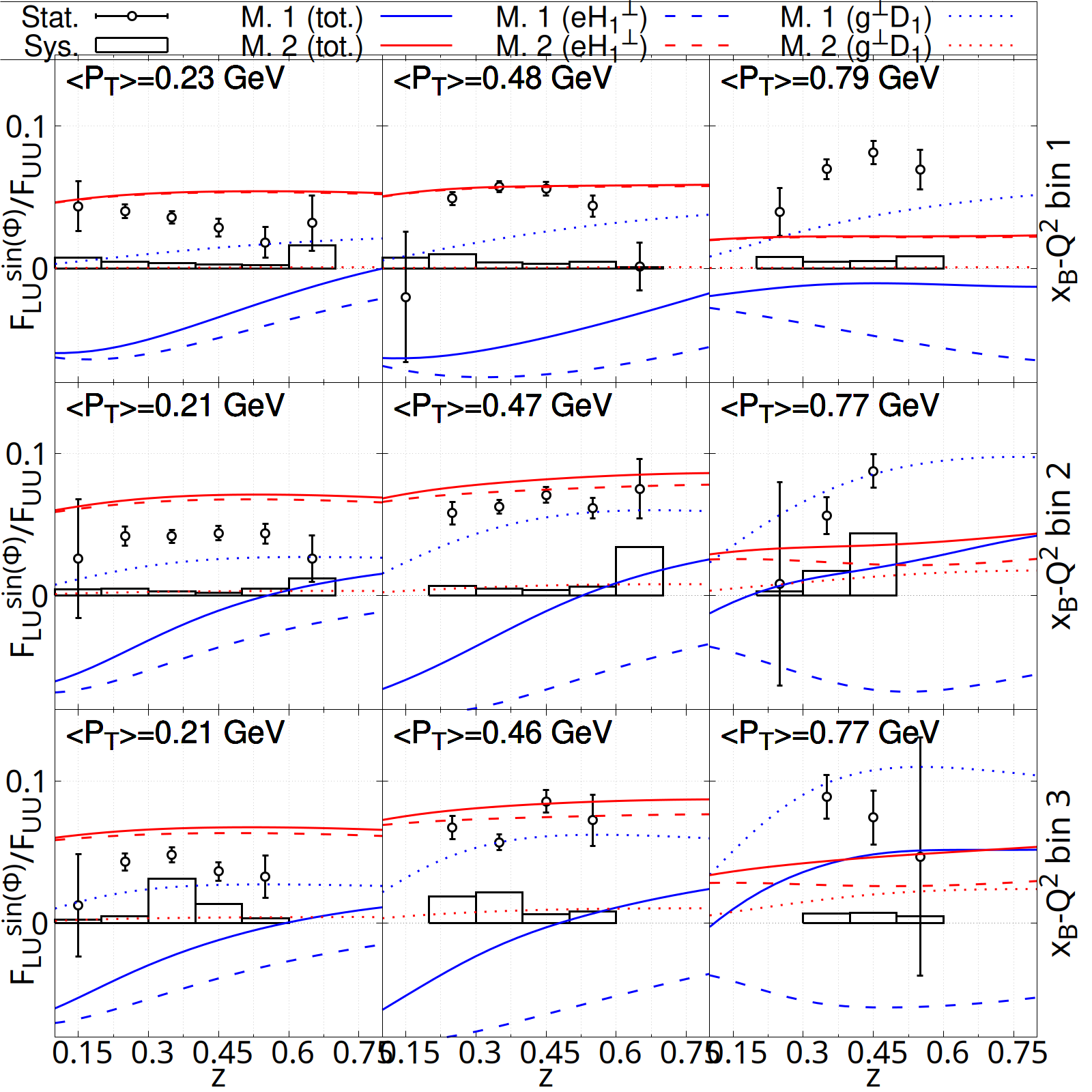}
	\caption{$z$ dependence of $F_{LU}^{\sin{\phi}}/F_{UU}$ for $K^+$ for increasing $P_T$ bins (left to right) and for increasing $Q^2-x_B$-bins (bin 1: $\langle Q^2\rangle=1.68$ GeV$^2$, $\langle x_B\rangle=0.15$, bin 2: $\langle Q^2\rangle=2.52$ GeV$^2$, $\langle x_B\rangle=0.26$, bin 3: $\langle Q^2\rangle=4.36$ GeV$^2$, $\langle x_B\rangle=0.37$ (The 3 rows correspond to the 3 $Q^2-x_B$-bins shown in Fig. \ref{fig:Q2_x_binning}.)). The systematic uncertainty is given by the black histogram. The predictions of the different theoretical models are shown by the bold and dashed lines (blue: model 1, red: model 2). More details about the models can be found in the text and in Refs. \cite{Mao2013,Mao2014}}
	\label{fig:Kp_z_4D}
	\end{center}
\end{figure*}

\begin{figure*}[h!]
\begin{center}
    \includegraphics[width=0.65\textwidth]{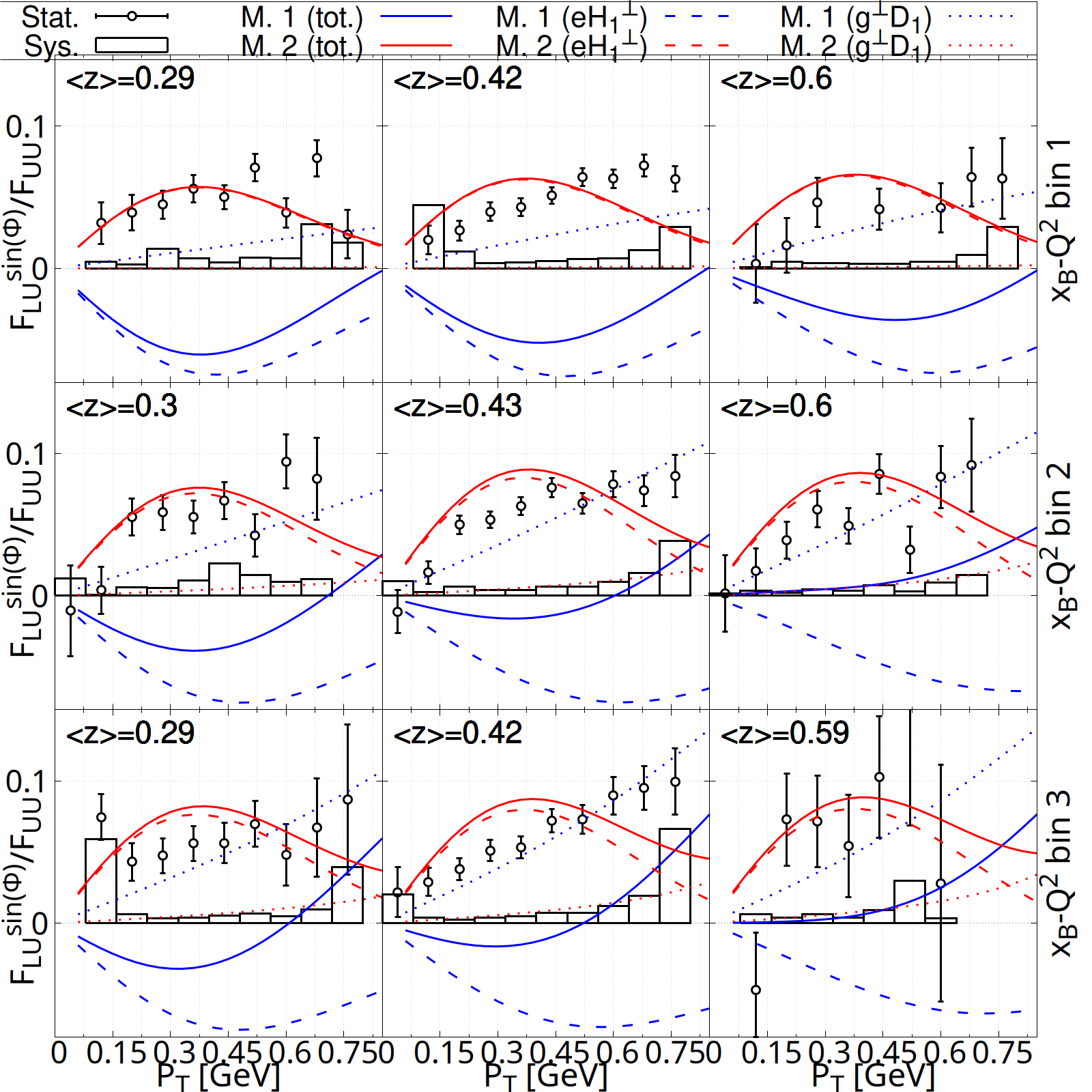}
	\caption{$P_T$ dependence of $F_{LU}^{\sin{\phi}}/F_{UU}$ for $K^+$ for increasing $z$ bins (left to right) and for increasing $Q^2-x_B$-bins (see caption of Fig. \ref{fig:Kp_z_4D}). The systematic uncertainty is given by the black histogram. The predictions of the different theoretical models are shown by the bold and dashed lines (see caption of Fig. \ref{fig:Kp_z_4D}).}
	\label{fig:Kp_PT_4D}
	\end{center}
\end{figure*}

Figures \ref{fig:Kp_z_4D} and \ref{fig:Kp_PT_4D} show the dependence on $z$ ($P_T$) for the $P_T$ ($z$) bins in different bins of $Q^2$ and $x_B$ for $eK^+X$. It can be observed that the general $P_T$ and $z$ behavior of the asymmetries is almost independent of $Q^2$ and $x_B$, and a rising trend in $P_T$ can be observed in most $z$ bins. For the $z$ dependence, the ratio is rising or constant up to $z$ = 0.4. For higher $z$ a falling trend may be observed in some bins.

\onecolumngrid

\twocolumngrid


The results are compared to theoretical predictions, calculated using the two models in Refs. \cite{Mao2013, Mao2014}, adjusted to kaons and to the chosen kinematic binning. 
Both models describe the proton as an active quark accompanied by spectator scalar and axial-vector diquarks; and both include the $eH_1^{\perp}$ and the $g^{\perp}D_1$ terms with the other terms assumed to be negligible. 
The first model uses a ratio for the axial-vector and scalar strengths fitted to data. 
In contrast to this, the second model uses a simpler propagator for the axial-vector diquark and the ratio of axial-vector and scalar is fixed by SU(4) spin-flavor symmetry. 
The FFs used in both models are described in Ref. \cite{ANSELMINO200998}. 
The models do not provide any uncertainties, only the mean value. 
Regarding Model 1, the $g^\perp D$ contribution alone bears some qualitative similarity to the data, but the associated $eH_1^{\perp}$ term is uniformly negative and this eliminates all agreement.
Turning to Model 2, one sees that the $P_T$ behavior of the data is semi-quantitatively reproduced.  
The $g D_1$ is typically small, but does contribute noticeably at the larger values of $P_T$, $z$.
Notwithstanding its contribution, Model 2 predicts an asymmetry that first increases with $P_T$ but then falls on $P_T\gtrsim 0.45$.
We have verified that physically constrained changes to the FFs do not materially affect these observations.

Furthermore, reviewing Figs.\,\ref{fig:Kp_z_4D}, \ref{fig:Kp_PT_4D}, one notes that the $eH_1^{\perp}$ and $g^\perp D_1$ contributions have the opposite sign in model 1 (blue dashed vs.\ blue dotted), 
whereas they have the same sign in model 2 (red dashed vs.\ red dotted).  
Since $H_1^{\perp}$, $D_1$ are unchanged for calculations in both models, then this behavior owes to different signs for $e$: it is negative in model 1 and positive in model 2 -- see, \textit{e.g}., Ref.\,\cite[Figs.\,1\,-\,3]{Mao2013}.


These kinematic and model dependencies can provide valuable insights into the underlying features of the involved TMDs and FFs. 
Although this is the first time that an analysis binned in all relevant kinematic variables was made available for positively charged kaons in the valence quark regime, Model 2 reproduces the sign and the tendencies in the data on $P_T \lesssim 0.7$ without any of its parameters varied, with the exception of the adjustments to kaons and to the measured kinematics. 
It follows that this approach is a good starting point for phenomenology. 
The fact that Model 2 may be challenged at higher $P_T$ could mean that the parametrizations of the involved TMDs and FFs should be improved or that additional terms from Eq. \ref{eq:FLUsin}, besides the two that have been used, provide measurable contributions within those kinematic regions.
Also higher-twist contributions, especially in the denominator of the structure function ratio, could contribute in some kinematic regions.
Another explanation may be that the factorization that was assumed is not valid at higher $P_T$. According to Ref. \cite{Bacchetta2022}, global fits of TMDs based on SIDIS and Drell-Yan data indicate that this may occur around $P_T$ $\approx$ 0.3-0.4 GeV, depending on the $Q^2$ and $z$ kinematic domain. Such effects can only be studied in a high-precision multidimensional investigation of kaon SIDIS, as it is available for the first time in the valence quark regime within this work.

In summary, the structure function ratio $F_{LU}^{\sin{\phi}}/F_{UU}$ corresponding to the polarized electron beam SSA in semi-inclusive deep inelastic scattering has been measured for the first time in the valence quark regime over a wide range of kinematics binned in all the relevant kinematic variables at the same time using positively charged kaons. At higher $P_T$ a drop in $F_{LU}^{\sin{\phi}}/F_{UU}$ may be observed for kaons in most bins. Comparison with theoretical predictions showed that the model describing the nucleon as an active quark and a spectator diquark, which has a single propagator, describes the sign and the tendencies in the data in most cases, and it predicts a falling behavior at high $P_T$. The data presented in this work will help to further constrain the TMDs and FFs in global fits.


We acknowledge the outstanding efforts of the staff of the Accelerator and the Physics Divisions at Jefferson Lab in making this experiment possible. 
This work was supported in part by the U.S. Department of Energy, the National Science Foundation (NSF), the Italian Istituto Nazionale di Fisica Nucleare (INFN), the French Centre National de la Recherche Scientifique (CNRS), the French Commissariat pour l$^{\prime}$Energie Atomique, the UK Science and Technology Facilities Council, the National Research Foundation (NRF) of Korea, the Helmholtz-Forschungsakademie Hessen f\"ur FAIR (HFHF), the Deutsche Forschungsgemeinschaft (DFG), the Chilean Agency of Research and Development (ANID), the National Natural Science Foundation of China (grant nos.\ 12135007, 12150013), Nanjing University of Posts and Telecommunications Science Foundation (grant no.\ NY223108), and the European Union’s Horizon 2020 Research and Innovation program under Grant Agreement No.\ 824093 (STRONG2020). This material is based upon work supported by the U.S. Department of Energy, Office of Science, Office of Nuclear Physics under contract DE-AC05-06OR23177.




\bibliography{kaon_sidis_paper}


\end{document}


\pagenumbering{gobble}

\section{Bin borders of the multidimensional binning}
The borders of the two dimensional cuts in $Q^2$ and $x_B$ and for each $Q^2$-$x_B$ bin, the bin borders in $z$ and $P_T$ are listed in the following tables.

\begin{table}[h!]
\begin{center}
\begin{tabular}{|l|l|}
\hline
Bin number & $Q^2$-$x_B$ bin\\ \hline
1          & $x_B<0.2$    \\ \hline
2          & $x_B>0.2$ and $Q^2<3.3$ GeV$^2$    \\ \hline
3          & $x_B>0.2$ and $Q^2>3.3$ GeV$^2$   \\ \hline
\end{tabular}
\end{center}
\caption{Bin borders and numbering of the bins on the $Q^2$-$x_B$-plane.}
  \label{tab:3D_Q2xbins}
\end{table}

\begin{table}[h!]
\begin{center}
\begin{tabular}{|l|l|l|}
\hline
Bin number & $P_T$ bin             & $z$ bin       \\ \hline
1          & $P_T<0.33$ GeV        & $z<0.34$      \\ \hline
2          & $0.33$ GeV$<P_T<0.66$ GeV & $0.34<z<0.52$ \\ \hline
3          & $P_T>0.66$ GeV        & $z>0.52$      \\ \hline
\end{tabular}
\end{center}
\caption{The binning in $z$ or in $P_T$.}
\label{tab:last_binning}
\end{table}

\section{Systematic uncertainties}
Absolute uncertainty on $F_{LU}^{\sin \phi}/F_{UU}$ from the different sources averaged over all bins of the fully multidimensional binning is shown in the following table.

\begin{table}[h!]
\centering
\begin{tabular}{|l|l|}
\hline
Source                       & Absolute uncertainty \\ \hline
Statistical uncertainty      & 0.02    \\ \hline
Pion contamination           & 0.005  \\ \hline
MC contamination             & 0.001   \\ \hline
Beam polarization            & 0.001  \\ \hline
Fiducial cuts                & 0.004 \\ \hline
Pion bin migration           & 0.001   \\ \hline
Radiative effects            & 0.002 \\ \hline
Acceptance                   & 0.002   \\ \hline
Cos terms                    & 0.002   \\ \hline
Bin migration                & 0.001   \\ \hline
Phi bin migration            & 0.0005  \\ \hline
Charge-symmetric background  & 0.0001  \\ \hline
Accidentals                  & 0.0005  \\ \hline
Total systematic uncertainty & 0.01   \\ \hline
\end{tabular}
\caption{Absolute uncertainty from the different sources averaged over all bins of the fully multidimensional binning}
    \label{tab:tot_4D_avg}
\end{table}

\newpage

\section{Summary table for the fully multidimensional study}
The following table shows the mean value of the kinematic variables, as well as the $F_{LU}^{\sin \phi}/F_{UU}$ and the statistical and systematic uncertainties for the fully differential study.

\begin{longtable}{|l|l|l|l|l|l|l|l|l|l|l|}
\hline
\textless{}$x_B$\textgreater{} & \adjustbox{stack=ll}{\textless{}$Q^2$\textgreater{}\\$[$GeV$^2]$} & \textless{}$z$\textgreater{} & \adjustbox{stack=ll}{\textless{}$P_T$\textgreater{}\\$[$GeV$]$} & \adjustbox{stack=ll}{\textless{}$W$\textgreater{}\\$[$GeV$]$} & \textless{}$y$\textgreater{} & \adjustbox{stack=ll}{\textless{}$M_X$\textgreater{}\\$[$GeV$]$} & \textless{}$\varepsilon$\textgreater{} & $F_{LU}^{\sin{\phi}}/F_{UU}$     & \multicolumn{1}{l|}{\adjustbox{stack=ll}{Stat.\\unc.}} & \adjustbox{stack=ll}{Sys.\\unc.} \\ \hline
\multicolumn{11}{|l|}{$x-Q^2$ bin 1} \\ \hline
\multicolumn{11}{|l|}{$P_T$ bin 1} \\ \hline
0.144 & 1.983 & 0.150 & 0.215 & 3.561 & 0.703 & 3.006 & 0.531 & 0.044 & 0.017 & 0.008 \\ \hline
0.152 & 1.920 & 0.250 & 0.238 & 3.413 & 0.649 & 2.805 & 0.606 & 0.040 & 0.005 & 0.005 \\ \hline
0.156 & 1.824 & 0.350 & 0.254 & 3.277 & 0.598 & 2.573 & 0.667 & 0.036 & 0.004 & 0.004 \\ \hline
0.160 & 1.710 & 0.450 & 0.273 & 3.143 & 0.549 & 2.320 & 0.720 & 0.029 & 0.006 & 0.003 \\ \hline
0.161 & 1.613 & 0.550 & 0.290 & 3.044 & 0.512 & 2.083 & 0.758 & 0.018 & 0.011 & 0.002 \\ \hline
0.160 & 1.599 & 0.650 & 0.303 & 3.046 & 0.511 & 1.901 & 0.763 & 0.032 & 0.020 & 0.016 \\ \hline
\multicolumn{11}{|l|}{$P_T$ bin 2} \\ \hline
0.121 & 1.725 & 0.150 & 0.364 & 3.661 & 0.721 & 3.025 & 0.506 & -0.020 & 0.046 & 0.008 \\ \hline
0.137 & 1.767 & 0.250 & 0.464 & 3.471 & 0.658 & 2.734 & 0.595 & 0.049 & 0.004 & 0.010 \\ \hline
0.143 & 1.697 & 0.350 & 0.499 & 3.328 & 0.606 & 2.503 & 0.659 & 0.057 & 0.004 & 0.004 \\ \hline
0.147 & 1.621 & 0.450 & 0.519 & 3.213 & 0.565 & 2.275 & 0.705 & 0.056 & 0.005 & 0.003 \\ \hline
0.149 & 1.573 & 0.550 & 0.536 & 3.150 & 0.542 & 2.067 & 0.730 & 0.044 & 0.007 & 0.005 \\ \hline
0.147 & 1.585 & 0.650 & 0.549 & 3.173 & 0.549 & 1.898 & 0.725 & 0.001 & 0.017 & 0.001 \\ \hline
\multicolumn{11}{|l|}{$P_T$ bin 3} \\ \hline
0.120 & 1.652 & 0.250 & 0.707 & 3.606 & 0.696 & 2.647 & 0.543 & 0.040 & 0.017 & 0.008 \\ \hline
0.134 & 1.679 & 0.350 & 0.768 & 3.435 & 0.640 & 2.376 & 0.619 & 0.070 & 0.007 & 0.005 \\ \hline
0.139 & 1.635 & 0.450 & 0.814 & 3.320 & 0.599 & 2.138 & 0.668 & 0.081 & 0.008 & 0.005 \\ \hline
0.140 & 1.623 & 0.550 & 0.833 & 3.297 & 0.591 & 1.973 & 0.678 & 0.069 & 0.014 & 0.008 \\ \hline
\multicolumn{11}{|l|}{$x-Q^2$ bin 2} \\ \hline
\multicolumn{11}{|l|}{$P_T$ bin 1} \\ \hline
0.220 & 2.972 & 0.150 & 0.186 & 3.380 & 0.691 & 2.840 & 0.544 & 0.026 & 0.042 & 0.005 \\ \hline
0.241 & 2.755 & 0.250 & 0.222 & 3.103 & 0.590 & 2.518 & 0.674 & 0.042 & 0.007 & 0.005 \\ \hline
0.262 & 2.581 & 0.350 & 0.234 & 2.872 & 0.511 & 2.226 & 0.759 & 0.042 & 0.005 & 0.003 \\ \hline
0.273 & 2.442 & 0.450 & 0.246 & 2.731 & 0.464 & 2.002 & 0.805 & 0.044 & 0.005 & 0.002 \\ \hline
0.268 & 2.371 & 0.550 & 0.258 & 2.718 & 0.455 & 1.864 & 0.815 & 0.044 & 0.007 & 0.005 \\ \hline
0.254 & 2.407 & 0.650 & 0.269 & 2.819 & 0.484 & 1.773 & 0.790 & 0.026 & 0.016 & 0.012 \\ \hline
\multicolumn{11}{|l|}{$P_T$ bin 2} \\ \hline
0.232 & 2.785 & 0.250 & 0.437 & 3.185 & 0.616 & 2.474 & 0.643 & 0.058 & 0.008 & 0.007 \\ \hline
0.251 & 2.587 & 0.350 & 0.476 & 2.941 & 0.531 & 2.164 & 0.739 & 0.063 & 0.005 & 0.005 \\ \hline
0.259 & 2.464 & 0.450 & 0.495 & 2.825 & 0.490 & 1.967 & 0.781 & 0.071 & 0.005 & 0.004 \\ \hline
0.255 & 2.428 & 0.550 & 0.509 & 2.828 & 0.488 & 1.846 & 0.785 & 0.062 & 0.007 & 0.006 \\ \hline
0.243 & 2.497 & 0.650 & 0.522 & 2.939 & 0.524 & 1.764 & 0.751 & 0.075 & 0.021 & 0.034 \\ \hline
\multicolumn{11}{|l|}{$P_T$ bin 3} \\ \hline
0.216 & 2.982 & 0.250 & 0.689 & 3.421 & 0.701 & 2.476 & 0.529 & 0.008 & 0.072 & 0.003 \\ \hline
0.232 & 2.738 & 0.350 & 0.740 & 3.154 & 0.602 & 2.124 & 0.660 & 0.056 & 0.013 & 0.017 \\ \hline
0.240 & 2.604 & 0.450 & 0.779 & 3.025 & 0.555 & 1.904 & 0.716 & 0.088 & 0.012 & 0.044 \\ \hline
\multicolumn{11}{|l|}{$x-Q^2$ bin 3} \\ \hline
\multicolumn{11}{|l|}{$P_T$ bin 1} \\ \hline
0.289 & 4.016 & 0.150 & 0.165 & 3.280 & 0.708 & 2.748 & 0.511 & 0.013 & 0.036 & 0.002 \\ \hline
0.357 & 4.478 & 0.250 & 0.214 & 2.998 & 0.645 & 2.421 & 0.594 & 0.043 & 0.006 & 0.005 \\ \hline
0.388 & 4.485 & 0.350 & 0.227 & 2.831 & 0.597 & 2.189 & 0.653 & 0.048 & 0.005 & 0.031 \\ \hline
0.396 & 4.404 & 0.450 & 0.237 & 2.764 & 0.573 & 2.023 & 0.681 & 0.036 & 0.007 & 0.014 \\ \hline
0.386 & 4.322 & 0.550 & 0.249 & 2.790 & 0.575 & 1.903 & 0.681 & 0.033 & 0.015 & 0.004 \\ \hline
\multicolumn{11}{|l|}{$P_T$ bin 2} \\ \hline
0.323 & 4.246 & 0.250 & 0.433 & 3.131 & 0.671 & 2.423 & 0.561 & 0.067 & 0.008 & 0.019 \\ \hline
0.363 & 4.379 & 0.350 & 0.474 & 2.936 & 0.621 & 2.158 & 0.626 & 0.057 & 0.005 & 0.022 \\ \hline
0.372 & 4.331 & 0.450 & 0.493 & 2.870 & 0.598 & 1.999 & 0.653 & 0.086 & 0.008 & 0.006 \\ \hline
0.362 & 4.269 & 0.550 & 0.509 & 2.902 & 0.604 & 1.889 & 0.648 & 0.072 & 0.018 & 0.008 \\ \hline
\multicolumn{11}{|l|}{$P_T$ bin 3} \\ \hline
0.317 & 4.153 & 0.350 & 0.740 & 3.136 & 0.668 & 2.105 & 0.567 & 0.089 & 0.015 & 0.007 \\ \hline
0.332 & 4.177 & 0.450 & 0.784 & 3.050 & 0.643 & 1.919 & 0.601 & 0.074 & 0.019 & 0.007 \\ \hline
0.323 & 4.139 & 0.550 & 0.808 & 3.092 & 0.653 & 1.829 & 0.587 & 0.047 & 0.084 & 0.005 \\ \hline
\hline
\multicolumn{11}{|l|}{$x-Q^2$ bin 1} \\ \hline
\multicolumn{11}{|l|}{$z$ bin 1} \\ \hline
0.159 & 1.937 & 0.320 & 0.120 & 3.341 & 0.627 & 2.695 & 0.634 & 0.032 & 0.015 & 0.004 \\ \hline
0.158 & 1.892 & 0.320 & 0.200 & 3.320 & 0.617 & 2.660 & 0.645 & 0.039 & 0.013 & 0.003 \\ \hline
0.152 & 1.804 & 0.320 & 0.280 & 3.304 & 0.605 & 2.622 & 0.660 & 0.045 & 0.010 & 0.014 \\ \hline
0.146 & 1.756 & 0.320 & 0.360 & 3.336 & 0.612 & 2.618 & 0.652 & 0.056 & 0.010 & 0.007 \\ \hline
0.143 & 1.723 & 0.320 & 0.440 & 3.355 & 0.616 & 2.593 & 0.646 & 0.050 & 0.008 & 0.004 \\ \hline
0.140 & 1.706 & 0.320 & 0.520 & 3.372 & 0.621 & 2.558 & 0.642 & 0.071 & 0.010 & 0.008 \\ \hline
0.138 & 1.702 & 0.321 & 0.600 & 3.402 & 0.630 & 2.527 & 0.631 & 0.039 & 0.010 & 0.007 \\ \hline
0.134 & 1.700 & 0.321 & 0.680 & 3.444 & 0.644 & 2.499 & 0.615 & 0.077 & 0.013 & 0.031 \\ \hline
0.130 & 1.707 & 0.322 & 0.760 & 3.508 & 0.665 & 2.484 & 0.586 & 0.024 & 0.017 & 0.018 \\ \hline
\multicolumn{11}{|l|}{$z$ bin 2} \\ \hline
0.163 & 1.843 & 0.419 & 0.120 & 3.217 & 0.581 & 2.442 & 0.684 & 0.020 & 0.010 & 0.045 \\ \hline
0.161 & 1.755 & 0.421 & 0.200 & 3.168 & 0.560 & 2.388 & 0.707 & 0.027 & 0.007 & 0.012 \\ \hline
0.156 & 1.693 & 0.422 & 0.280 & 3.167 & 0.555 & 2.365 & 0.715 & 0.040 & 0.006 & 0.004 \\ \hline
0.151 & 1.661 & 0.421 & 0.360 & 3.204 & 0.564 & 2.368 & 0.705 & 0.043 & 0.006 & 0.004 \\ \hline
0.147 & 1.643 & 0.421 & 0.440 & 3.234 & 0.573 & 2.358 & 0.695 & 0.051 & 0.006 & 0.005 \\ \hline
0.145 & 1.635 & 0.421 & 0.520 & 3.252 & 0.578 & 2.330 & 0.690 & 0.064 & 0.006 & 0.007 \\ \hline
0.143 & 1.632 & 0.423 & 0.600 & 3.272 & 0.584 & 2.295 & 0.684 & 0.063 & 0.006 & 0.007 \\ \hline
0.141 & 1.634 & 0.425 & 0.680 & 3.294 & 0.591 & 2.253 & 0.677 & 0.072 & 0.008 & 0.013 \\ \hline
0.139 & 1.640 & 0.427 & 0.760 & 3.325 & 0.601 & 2.212 & 0.666 & 0.063 & 0.009 & 0.029 \\ \hline
\multicolumn{11}{|l|}{$z$ bin 3} \\ \hline
0.166 & 1.655 & 0.598 & 0.120 & 3.023 & 0.509 & 1.993 & 0.762 & 0.003 & 0.028 & 0.001 \\ \hline
0.162 & 1.598 & 0.595 & 0.200 & 3.021 & 0.504 & 1.985 & 0.768 & 0.016 & 0.019 & 0.004 \\ \hline
0.157 & 1.579 & 0.593 & 0.280 & 3.054 & 0.512 & 1.992 & 0.760 & 0.046 & 0.017 & 0.004 \\ \hline
0.149 & 1.569 & 0.591 & 0.440 & 3.138 & 0.538 & 1.999 & 0.735 & 0.042 & 0.015 & 0.003 \\ \hline
0.145 & 1.587 & 0.590 & 0.600 & 3.206 & 0.560 & 1.970 & 0.713 & 0.043 & 0.017 & 0.005 \\ \hline
0.143 & 1.604 & 0.590 & 0.680 & 3.245 & 0.573 & 1.950 & 0.699 & 0.064 & 0.021 & 0.009 \\ \hline
0.141 & 1.625 & 0.590 & 0.760 & 3.289 & 0.588 & 1.928 & 0.682 & 0.063 & 0.028 & 0.029 \\ \hline
\multicolumn{11}{|l|}{$x-Q^2$ bin 2} \\ \hline
\multicolumn{11}{|l|}{$z$ bin 1} \\ \hline
0.256 & 2.689 & 0.321 & 0.040 & 2.966 & 0.546 & 2.378 & 0.724 & -0.010 & 0.032 & 0.012 \\ \hline
0.256 & 2.640 & 0.321 & 0.120 & 2.940 & 0.535 & 2.347 & 0.735 & 0.004 & 0.016 & 0.001 \\ \hline
0.256 & 2.627 & 0.321 & 0.200 & 2.933 & 0.532 & 2.321 & 0.739 & 0.055 & 0.013 & 0.006 \\ \hline
0.254 & 2.636 & 0.321 & 0.280 & 2.950 & 0.536 & 2.308 & 0.734 & 0.059 & 0.012 & 0.005 \\ \hline
0.252 & 2.628 & 0.321 & 0.360 & 2.962 & 0.540 & 2.280 & 0.730 & 0.055 & 0.012 & 0.011 \\ \hline
0.247 & 2.622 & 0.321 & 0.440 & 2.988 & 0.547 & 2.255 & 0.724 & 0.067 & 0.013 & 0.023 \\ \hline
0.240 & 2.664 & 0.322 & 0.520 & 3.054 & 0.568 & 2.260 & 0.701 & 0.042 & 0.015 & 0.015 \\ \hline
0.234 & 2.724 & 0.322 & 0.600 & 3.135 & 0.596 & 2.271 & 0.670 & 0.094 & 0.019 & 0.010 \\ \hline
0.227 & 2.808 & 0.323 & 0.680 & 3.233 & 0.630 & 2.292 & 0.627 & 0.082 & 0.029 & 0.012 \\ \hline
\multicolumn{11}{|l|}{$z$ bin 2} \\ \hline
0.276 & 2.525 & 0.434 & 0.040 & 2.760 & 0.477 & 2.074 & 0.791 & -0.011 & 0.015 & 0.010 \\ \hline
0.273 & 2.489 & 0.430 & 0.120 & 2.756 & 0.474 & 2.067 & 0.795 & 0.016 & 0.008 & 0.003 \\ \hline
0.271 & 2.463 & 0.429 & 0.200 & 2.756 & 0.472 & 2.052 & 0.797 & 0.050 & 0.007 & 0.006 \\ \hline
0.268 & 2.463 & 0.429 & 0.280 & 2.768 & 0.475 & 2.037 & 0.794 & 0.053 & 0.006 & 0.004 \\ \hline
0.264 & 2.470 & 0.428 & 0.360 & 2.795 & 0.483 & 2.027 & 0.787 & 0.063 & 0.006 & 0.004 \\ \hline
0.259 & 2.479 & 0.427 & 0.440 & 2.830 & 0.493 & 2.016 & 0.778 & 0.076 & 0.007 & 0.006 \\ \hline
0.254 & 2.498 & 0.427 & 0.520 & 2.872 & 0.505 & 2.001 & 0.767 & 0.065 & 0.007 & 0.007 \\ \hline
0.249 & 2.530 & 0.429 & 0.600 & 2.924 & 0.522 & 1.986 & 0.751 & 0.078 & 0.009 & 0.010 \\ \hline
0.243 & 2.572 & 0.432 & 0.680 & 2.982 & 0.541 & 1.969 & 0.732 & 0.074 & 0.011 & 0.016 \\ \hline
0.239 & 2.621 & 0.436 & 0.760 & 3.042 & 0.561 & 1.946 & 0.710 & 0.084 & 0.015 & 0.039 \\ \hline
\multicolumn{11}{|l|}{$z$ bin 3} \\ \hline
0.266 & 2.364 & 0.589 & 0.040 & 2.724 & 0.456 & 1.821 & 0.815 & 0.002 & 0.027 & 0.002 \\ \hline
0.265 & 2.380 & 0.588 & 0.120 & 2.738 & 0.460 & 1.825 & 0.811 & 0.017 & 0.016 & 0.004 \\ \hline
0.262 & 2.382 & 0.585 & 0.200 & 2.754 & 0.465 & 1.826 & 0.806 & 0.039 & 0.013 & 0.002 \\ \hline
0.259 & 2.385 & 0.584 & 0.280 & 2.774 & 0.471 & 1.822 & 0.802 & 0.061 & 0.013 & 0.005 \\ \hline
0.256 & 2.407 & 0.583 & 0.360 & 2.806 & 0.481 & 1.818 & 0.792 & 0.049 & 0.012 & 0.003 \\ \hline
0.252 & 2.435 & 0.583 & 0.440 & 2.847 & 0.494 & 1.814 & 0.780 & 0.085 & 0.014 & 0.007 \\ \hline
0.247 & 2.469 & 0.582 & 0.520 & 2.896 & 0.510 & 1.809 & 0.765 & 0.032 & 0.016 & 0.003 \\ \hline
0.243 & 2.508 & 0.582 & 0.600 & 2.949 & 0.527 & 1.800 & 0.748 & 0.084 & 0.022 & 0.009 \\ \hline
0.238 & 2.563 & 0.582 & 0.680 & 3.010 & 0.548 & 1.792 & 0.726 & 0.092 & 0.033 & 0.015 \\ \hline
\multicolumn{11}{|l|}{$x-Q^2$ bin 3} \\ \hline
\multicolumn{11}{|l|}{$z$ bin 1} \\ \hline
0.385 & 4.527 & 0.320 & 0.120 & 2.856 & 0.607 & 2.268 & 0.641 & 0.075 & 0.016 & 0.059  \\ \hline
0.384 & 4.508 & 0.320 & 0.200 & 2.855 & 0.605 & 2.248 & 0.643 & 0.043 & 0.013 & 0.006 \\ \hline
0.380 & 4.493 & 0.321 & 0.280 & 2.876 & 0.610 & 2.237 & 0.638 & 0.047 & 0.012 & 0.004 \\ \hline
0.370 & 4.452 & 0.321 & 0.360 & 2.916 & 0.619 & 2.234 & 0.627 & 0.056 & 0.012 & 0.004 \\ \hline
0.359 & 4.403 & 0.321 & 0.440 & 2.964 & 0.630 & 2.230 & 0.615 & 0.056 & 0.014 & 0.005 \\ \hline
0.344 & 4.333 & 0.321 & 0.520 & 3.025 & 0.644 & 2.230 & 0.598 & 0.070 & 0.016 & 0.007 \\ \hline
0.326 & 4.251 & 0.322 & 0.600 & 3.107 & 0.664 & 2.242 & 0.573 & 0.048 & 0.021 & 0.005 \\ \hline
0.306 & 4.113 & 0.323 & 0.680 & 3.196 & 0.684 & 2.255 & 0.545 & 0.067 & 0.035 & 0.010 \\ \hline
0.285 & 3.961 & 0.325 & 0.760 & 3.290 & 0.706 & 2.267 & 0.516 & 0.087 & 0.053 & 0.040 \\ \hline
\multicolumn{11}{|l|}{$z$ bin 2} \\ \hline
0.402 & 4.456 & 0.421 & 0.040 & 2.752 & 0.573 & 2.073 & 0.681 & 0.022 & 0.018 & 0.020 \\ \hline
0.398 & 4.445 & 0.420 & 0.120 & 2.764 & 0.576 & 2.076 & 0.678 & 0.029 & 0.010 & 0.004 \\ \hline
0.395 & 4.435 & 0.419 & 0.200 & 2.777 & 0.579 & 2.071 & 0.675 & 0.038 & 0.008 & 0.002 \\ \hline
0.390 & 4.411 & 0.419 & 0.280 & 2.799 & 0.583 & 2.065 & 0.670 & 0.051 & 0.007 & 0.004 \\ \hline
0.383 & 4.395 & 0.419 & 0.360 & 2.830 & 0.591 & 2.057 & 0.661 & 0.053 & 0.008 & 0.005 \\ \hline
0.374 & 4.361 & 0.419 & 0.440 & 2.869 & 0.600 & 2.048 & 0.651 & 0.072 & 0.008 & 0.007 \\ \hline
0.364 & 4.327 & 0.420 & 0.520 & 2.909 & 0.609 & 2.033 & 0.640 & 0.073 & 0.010 & 0.007 \\ \hline
0.354 & 4.287 & 0.421 & 0.600 & 2.957 & 0.621 & 2.015 & 0.627 & 0.090 & 0.013 & 0.012 \\ \hline
0.342 & 4.241 & 0.424 & 0.680 & 3.010 & 0.634 & 1.995 & 0.611 & 0.095 & 0.015 & 0.019 \\ \hline
0.330 & 4.190 & 0.428 & 0.760 & 3.063 & 0.647 & 1.966 & 0.595 & 0.099 & 0.023 & 0.066 \\ \hline
\multicolumn{11}{|l|}{$z$ bin 3} \\ \hline
0.379 & 4.290 & 0.585 & 0.120 & 2.814 & 0.580 & 1.866 & 0.676 & -0.047 & 0.040 & 0.007 \\ \hline
0.377 & 4.300 & 0.584 & 0.200 & 2.827 & 0.584 & 1.863 & 0.671 & 0.073 & 0.032 & 0.004 \\ \hline
0.372 & 4.294 & 0.584 & 0.280 & 2.852 & 0.591 & 1.861 & 0.664 & 0.072 & 0.032 & 0.006 \\ \hline
0.365 & 4.277 & 0.584 & 0.360 & 2.885 & 0.599 & 1.859 & 0.654 & 0.054 & 0.036 & 0.004 \\ \hline
0.357 & 4.256 & 0.583 & 0.440 & 2.923 & 0.609 & 1.854 & 0.642 & 0.103 & 0.043 & 0.009 \\ \hline
0.349 & 4.226 & 0.584 & 0.520 & 2.964 & 0.620 & 1.844 & 0.630 & 0.125 & 0.056 & 0.030 \\ \hline
0.339 & 4.195 & 0.583 & 0.600 & 3.014 & 0.632 & 1.835 & 0.614 & 0.028 & 0.083 & 0.004 \\ \hline
\caption{Result of the 4-dimensional study with $z$ or $P_T$ as the main bin variable with 3 $x_B-Q^2$ bins and 3 $P_T$ or $z$ bins. The columns \textless{}$x_B$\textgreater{}, \textless{}$Q^2$\textgreater{} , \textless{}$z$\textgreater{}, \textless{}$P_T$\textgreater{}, \textless{}$W$\textgreater{}, \textless{}$y$\textgreater{}, \textless{}$M_X$\textgreater{} and \textless{}$\varepsilon$\textgreater{} provide the mean value of the stated variable or the bin middle if it is the main bin variable. The columns Stat. and Sys. provide the statistical and systematic uncertainties of $F_{LU}^{\sin{\phi}}/F_{UU}$.}
\end{longtable}

\newpage

\section{Deep neural network used  to improve PID}

The root TMVA was chosen as a machine learning environment for building, training, and evaluating the deep neural network (DNN) used to lower the pion contamination in the kaon sample. The neural network consists of 3 fully connected hidden layers with 128 neurons per layer. Each neuron uses the tanh activation function, except the last one, which is linear. The weights were initialized using the Xavier initialization method \cite{pmlr-v9-glorot10a}. They were optimized using the ADAM optimizer \cite{kingma2017adammethodstochasticoptimization} with cross-entropy loss function, with a batch size of 30 and with a learning rate of 10$^{-5}$. The input variables were normalized to 0-1 interval. To prevent over-fitting, the training was stopped after 50 steps, after both the training and the validation uncertainties started to rise. The resulting Receiver Operating Characteristics (ROC) curve for different momenta ranges and the DNN response are shown below. The following quantities are used as input:
\begin{itemize}
    \item Momentum, ToF $\beta$ and $\chi^2$ of the PID
    \item Time information ($\beta$) from all ToF, DC, and ECAL layers
    \item HTCC number of photo-electrons
    \item Energy deposition and shower profile in the calorimeter layers
\end{itemize}

\begin{figure}[h!]
\begin{center}
    \includegraphics[width=0.65\textwidth]{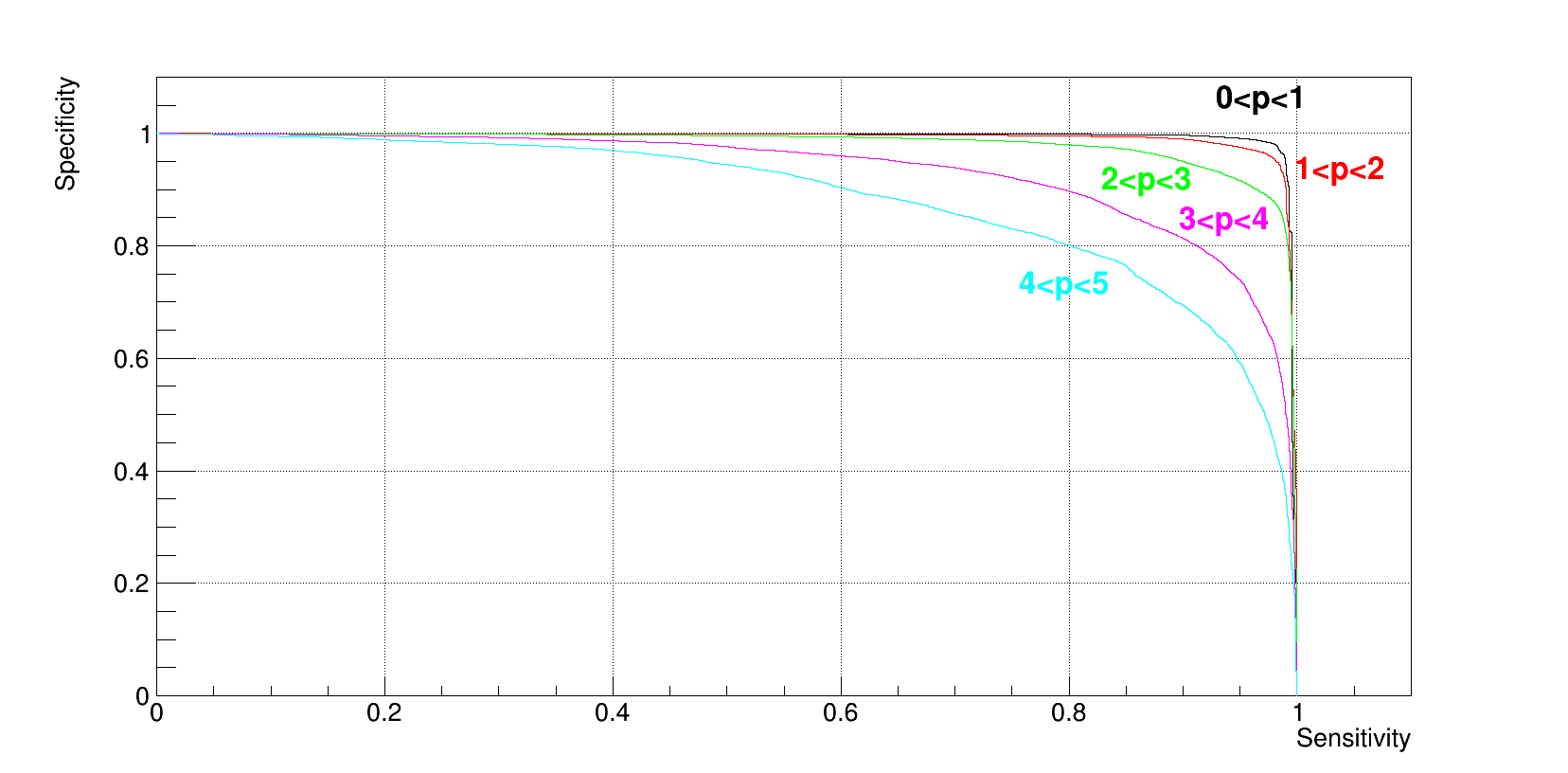}
	\caption{ROC curves shown at different momenta [GeV] ranges.}
	\label{fig:Q2_x_binning}
	\end{center}
\end{figure}

\begin{figure}[h!]
\begin{center}
    \includegraphics[width=0.65\textwidth]{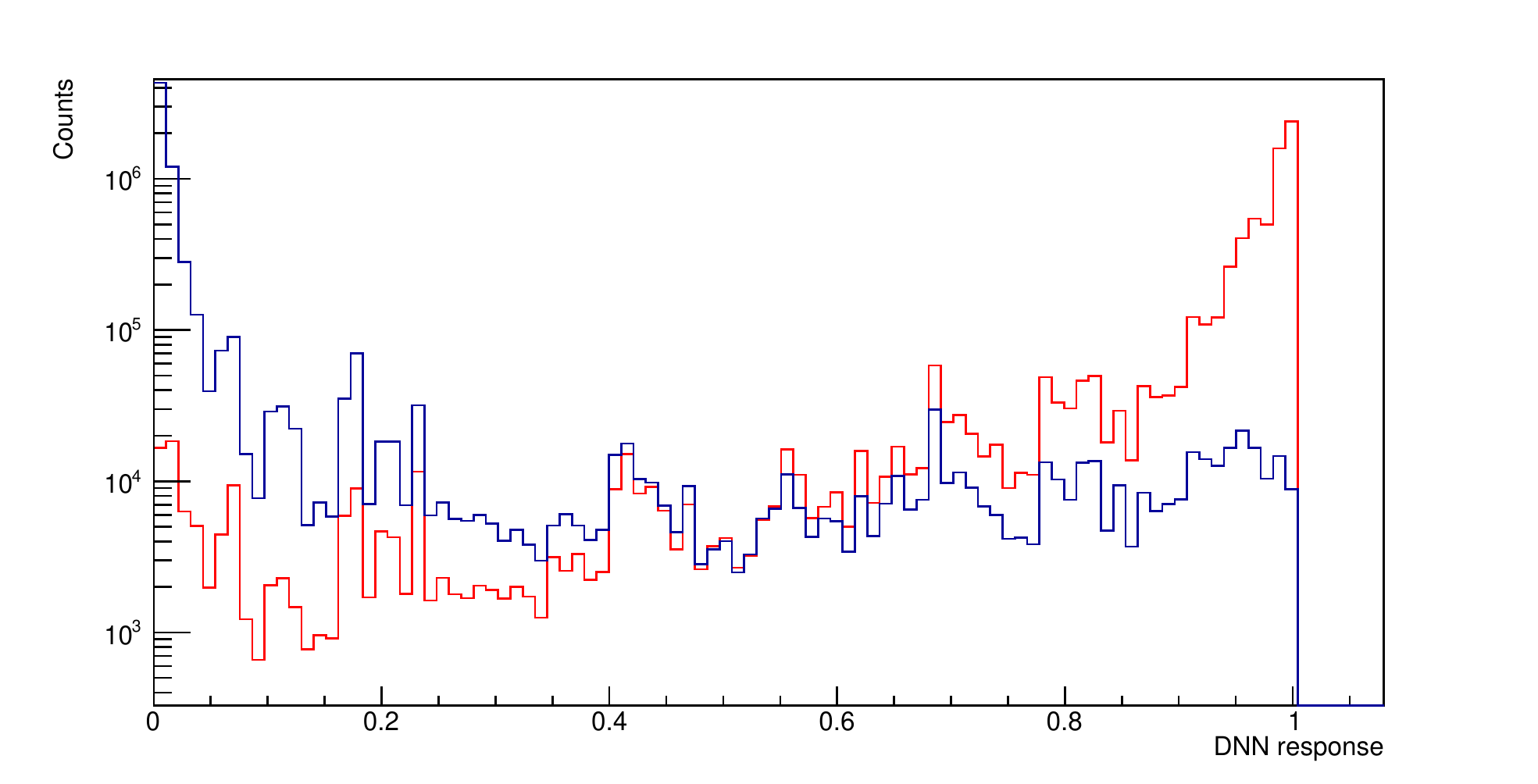}
	\caption{DNN response for the whole momentum range. The kaons are shown in red and the pions in blue. A subset of the training sample is shown here.}
	\label{fig:Q2_x_binning}
	\end{center}
\end{figure}

\newpage

\section{Details of the correction of the asymmetries due to pion contamination}
The following table shows the corrected $F_{LU}^{\sin{\phi}}/F_{UU}$, the raw extracted $F_{LU}^{\sin{\phi}}/F_{UU}$, the extracted pion $F_{LU}^{\sin{\phi}}/F_{UU}$, and the contamination estimates based on Monte Carlo simulations, as well as their statistical uncertainties for the fully differential study. The pion values stated here were used for the subtraction in the kaon sample. These values are consistent with the previously published CLAS12 pion values \cite{PhysRevLett.128.062005}, as they were merely re-binned so that they match the binning used for kaons.

\begin{longtable}{|l|l|l|l|l|l|l|l|l|} 
\hline
$z$/$P_T$ [GeV] & \multicolumn{1}{l|}{\adjustbox{stack=ll}{Corrected\\$F_{LU}^{\sin{\phi}}/F_{UU}$}}     & \multicolumn{1}{l|}{\adjustbox{stack=ll}{Unc. of Corrected\\$F_{LU}^{\sin{\phi}}/F_{UU}$}} & \multicolumn{1}{l|}{\adjustbox{stack=ll}{Raw\\$F_{LU}^{\sin{\phi}}/F_{UU}$}}     & \multicolumn{1}{l|}{\adjustbox{stack=ll}{Unc. of Raw\\$F_{LU}^{\sin{\phi}}/F_{UU}$}} & \multicolumn{1}{l|}{\adjustbox{stack=ll}{Pion\\$F_{LU}^{\sin{\phi}}/F_{UU}$}}     & \multicolumn{1}{l|}{\adjustbox{stack=ll}{Unc. of Pion\\$F_{LU}^{\sin{\phi}}/F_{UU}$}} & Contamination & \multicolumn{1}{l|}{\adjustbox{stack=ll}{Unc. of\\Contamination}}   \\ \hline
\multicolumn{9}{|l|}{$x-Q^2$ bin 1}                                                                                                                                                                                                                                                                                                                                                                                                                                                      \\ \hline
\multicolumn{9}{|l|}{$P_T$ bin 1}                                                                                                                                                                                                                                                                                                                                                                                                                                                        \\ 
\hline
0.1500 & 0.0437 & 0.0174 & 0.0411 & 0.0163 & 0.0031 & 0.0022 & 0.0628 & 0.0019 \\ \hline
0.2500 & 0.0401 & 0.0049 & 0.0377 & 0.0046 & 0.0044 & 0.0010 & 0.0672 & 0.0008 \\ \hline
0.3500 & 0.0359 & 0.0044 & 0.0343 & 0.0040 & 0.0156 & 0.0016 & 0.0789 & 0.0011 \\ \hline
0.4500 & 0.0286 & 0.0061 & 0.0289 & 0.0057 & 0.0330 & 0.0023 & 0.0673 & 0.0014 \\ \hline
0.5500 & 0.0183 & 0.0107 & 0.0204 & 0.0100 & 0.0513 & 0.0035 & 0.0612 & 0.0023 \\ \hline
0.6500 & 0.0318 & 0.0196 & 0.0343 & 0.0181 & 0.0628 & 0.0046 & 0.0798 & 0.0059 \\ \hline
\multicolumn{9}{|l|}{$P_T$ bin 2}                                                                                                                                                                                                                                                                                                                                                                                                                                                        \\ 
\hline
0.1500 & -0.0201 & 0.0459 & -0.0182 & 0.0431 & 0.0128 & 0.0045 & 0.0601 & 0.0050 \\ \hline
0.2500 & 0.0491 & 0.0044 & 0.0467 & 0.0041 & 0.0145 & 0.0010 & 0.0692 & 0.0007 \\ \hline
0.3500 & 0.0573 & 0.0038 & 0.0541 & 0.0035 & 0.0179 & 0.0011 & 0.0823 & 0.0007 \\ \hline
0.4500 & 0.0560 & 0.0050 & 0.0532 & 0.0046 & 0.0216 & 0.0014 & 0.0806 & 0.0009 \\ \hline
0.5500 & 0.0438 & 0.0073 & 0.0429 & 0.0066 & 0.0343 & 0.0021 & 0.0855 & 0.0015 \\ \hline
0.6500 & 0.0014 & 0.0167 & 0.0062 & 0.0151 & 0.0524 & 0.0032 & 0.0937 & 0.0042 \\ \hline
\multicolumn{9}{|l|}{$P_T$ bin 3}                                                                                                                                                                                                                                                                                                                                                                                                                                                        \\ 
\hline
0.2500 & 0.0397 & 0.0167 & 0.0375 & 0.0155 & 0.0110 & 0.0045 & 0.0762 & 0.0033 \\ \hline
0.3500 & 0.0697 & 0.0068 & 0.0654 & 0.0062 & 0.0231 & 0.0020 & 0.0919 & 0.0013 \\ \hline
0.4500 & 0.0814 & 0.0080 & 0.0756 & 0.0071 & 0.0275 & 0.0021 & 0.1077 & 0.0017 \\ \hline
0.5500 & 0.0694 & 0.0141 & 0.0637 & 0.0122 & 0.0263 & 0.0025 & 0.1331 & 0.0037 \\ \hline
\multicolumn{9}{|l|}{$x-Q^2$ bin 2}                                                                                                                                                                                                                                                                                                                                                                                                                                                      \\ \hline
\multicolumn{9}{|l|}{$P_T$ bin 1}                                                                                                                                                                                                                                                                                                                                                                                                                                                        \\ 
\hline
0.1500 & 0.0261 & 0.0416 & 0.0253 & 0.0390 & 0.0132 & 0.0049 & 0.0638 & 0.0044 \\ \hline
0.2500 & 0.0420 & 0.0067 & 0.0393 & 0.0062 & 0.0086 & 0.0013 & 0.0796 & 0.0012 \\ \hline
0.3500 & 0.0418 & 0.0046 & 0.0398 & 0.0042 & 0.0183 & 0.0014 & 0.0827 & 0.0009 \\ \hline
0.4500 & 0.0441 & 0.0051 & 0.0430 & 0.0046 & 0.0307 & 0.0018 & 0.0870 & 0.0009 \\ \hline
0.5500 & 0.0437 & 0.0071 & 0.0442 & 0.0064 & 0.0487 & 0.0028 & 0.0977 & 0.0015 \\ \hline
0.6500 & 0.0262 & 0.0162 & 0.0302 & 0.0145 & 0.0643 & 0.0041 & 0.1050 & 0.0040 \\ \hline
\multicolumn{9}{|l|}{$P_T$ bin 2}                                                                                                                                                                                                                                                                                                                                                                                                                                                        \\ 
\hline
0.2500 & 0.0581 & 0.0080 & 0.0553 & 0.0074 & 0.0233 & 0.0018 & 0.0814 & 0.0013 \\ \hline
0.3500 & 0.0626 & 0.0049 & 0.0593 & 0.0045 & 0.0292 & 0.0015 & 0.0984 & 0.0009 \\ \hline
0.4500 & 0.0709 & 0.0054 & 0.0672 & 0.0048 & 0.0371 & 0.0019 & 0.1095 & 0.0009 \\ \hline
0.5500 & 0.0616 & 0.0073 & 0.0609 & 0.0063 & 0.0564 & 0.0028 & 0.1372 & 0.0016 \\ \hline
0.6500 & 0.0751 & 0.0209 & 0.0749 & 0.0177 & 0.0737 & 0.0039 & 0.1518 & 0.0049 \\ \hline
\multicolumn{9}{|l|}{$P_T$ bin 3}                                                                                                                                                                                                                                                                                                                                                                                                                                                        \\ 
\hline
0.2500 & 0.0084 & 0.0716 & 0.0101 & 0.0656 & 0.0289 & 0.0165 & 0.0827 & 0.0133 \\ \hline
0.3500 & 0.0563 & 0.0130 & 0.0542 & 0.0117 & 0.0363 & 0.0038 & 0.1032 & 0.0028 \\ \hline
0.4500 & 0.0877 & 0.0116 & 0.0816 & 0.0098 & 0.0485 & 0.0035 & 0.1555 & 0.0030 \\ \hline
\multicolumn{9}{|l|}{$x-Q^2$ bin 3}                                                                                                                                                                                                                                                                                                                                                                                                                                                      \\ \hline
\multicolumn{9}{|l|}{$P_T$ bin 1}                                                                                                                                                                                                                                                                                                                                                                                                                                                        \\ 
\hline
0.1500 & 0.0126 & 0.0360 & 0.0125 & 0.0336 & 0.0111 & 0.0043 & 0.0684 & 0.0042 \\ \hline
0.2500 & 0.0431 & 0.0060 & 0.0402 & 0.0055 & 0.0130 & 0.0014 & 0.0955 & 0.0013 \\ \hline
0.3500 & 0.0481 & 0.0052 & 0.0453 & 0.0046 & 0.0229 & 0.0017 & 0.1125 & 0.0014 \\ \hline
0.4500 & 0.0363 & 0.0067 & 0.0360 & 0.0058 & 0.0336 & 0.0024 & 0.1301 & 0.0021 \\ \hline
0.5500 & 0.0328 & 0.0149 & 0.0345 & 0.0126 & 0.0437 & 0.0033 & 0.1497 & 0.0046 \\ \hline
\multicolumn{9}{|l|}{$P_T$ bin 2}                                                                                                                                                                                                                                                                                                                                                                                                                                                        \\ 
\hline
0.2500 & 0.0672 & 0.0083 & 0.0640 & 0.0075 & 0.0341 & 0.0022 & 0.0961 & 0.0016 \\ \hline
0.3500 & 0.0568 & 0.0054 & 0.0549 & 0.0047 & 0.0425 & 0.0020 & 0.1317 & 0.0014 \\ \hline
0.4500 & 0.0856 & 0.0078 & 0.0803 & 0.0065 & 0.0537 & 0.0026 & 0.1641 & 0.0021 \\ \hline
0.5500 & 0.0724 & 0.0180 & 0.0710 & 0.0143 & 0.0657 & 0.0034 & 0.2043 & 0.0056 \\ \hline
\multicolumn{9}{|l|}{$P_T$ bin 3}                                                                                                                                                                                                                                                                                                                                                                                                                                                        \\ 
\hline
0.3500 & 0.0887 & 0.0153 & 0.0825 & 0.0131 & 0.0448 & 0.0045 & 0.1417 & 0.0039 \\ \hline
0.4500 & 0.0742 & 0.0188 & 0.0715 & 0.0146 & 0.0617 & 0.0043 & 0.2179 & 0.0066 \\ \hline
0.5500 & 0.0468 & 0.0835 & 0.0547 & 0.0599 & 0.0748 & 0.0055 & 0.2821 & 0.0288 \\ \hline
\hline
\multicolumn{9}{|l|}{$x-Q^2$ bin 1}                                                                                                                                                                                                                                                                                                                                                                                                                                                       \\ 
\hline
\multicolumn{9}{|l|}{$z$ bin 1}                                                                                                                                                                                                                                                                                                                                                                                                                                                       \\ 
\hline
0.1200 & 0.0320 & 0.0146 & 0.0304 & 0.0134 & 0.0124 & 0.0049 & 0.0825 & 0.0043 \\ \hline
0.2000 & 0.0394 & 0.0125 & 0.0375 & 0.0114 & 0.0174 & 0.0036 & 0.0848 & 0.0032 \\ \hline
0.2800 & 0.0448 & 0.0099 & 0.0426 & 0.0091 & 0.0149 & 0.0031 & 0.0742 & 0.0024 \\ \hline
0.3600 & 0.0559 & 0.0097 & 0.0528 & 0.0089 & 0.0150 & 0.0027 & 0.0750 & 0.0020 \\ \hline
0.4400 & 0.0501 & 0.0085 & 0.0473 & 0.0078 & 0.0139 & 0.0024 & 0.0783 & 0.0018 \\ \hline
0.5200 & 0.0708 & 0.0097 & 0.0668 & 0.0089 & 0.0227 & 0.0026 & 0.0834 & 0.0019 \\ \hline
0.6000 & 0.0390 & 0.0101 & 0.0377 & 0.0093 & 0.0229 & 0.0031 & 0.0804 & 0.0021 \\ \hline
0.6800 & 0.0774 & 0.0128 & 0.0735 & 0.0118 & 0.0281 & 0.0038 & 0.0779 & 0.0025 \\ \hline
0.7600 & 0.0240 & 0.0171 & 0.0240 & 0.0156 & 0.0233 & 0.0053 & 0.0849 & 0.0036 \\ \hline
\multicolumn{9}{|l|}{$z$ bin 2}                                                                                                                                                                                                                                                                                                                                                                                                                                                       \\ 
\hline
0.1200 & 0.0202 & 0.0101 & 0.0202 & 0.0093 & 0.0203 & 0.0036 & 0.0761 & 0.0030 \\ \hline
0.2000 & 0.0266 & 0.0070 & 0.0264 & 0.0065 & 0.0245 & 0.0027 & 0.0715 & 0.0019 \\ \hline
0.2800 & 0.0398 & 0.0064 & 0.0391 & 0.0059 & 0.0293 & 0.0023 & 0.0687 & 0.0014 \\ \hline
0.3600 & 0.0431 & 0.0061 & 0.0417 & 0.0056 & 0.0242 & 0.0020 & 0.0738 & 0.0012 \\ \hline
0.4400 & 0.0513 & 0.0059 & 0.0487 & 0.0054 & 0.0192 & 0.0017 & 0.0793 & 0.0011 \\ \hline
0.5200 & 0.0640 & 0.0062 & 0.0602 & 0.0056 & 0.0190 & 0.0018 & 0.0860 & 0.0012 \\ \hline
0.6000 & 0.0630 & 0.0063 & 0.0594 & 0.0058 & 0.0221 & 0.0019 & 0.0889 & 0.0013 \\ \hline
0.6800 & 0.0723 & 0.0076 & 0.0674 & 0.0069 & 0.0197 & 0.0021 & 0.0935 & 0.0016 \\ \hline
0.7600 & 0.0629 & 0.0089 & 0.0589 & 0.0080 & 0.0233 & 0.0025 & 0.1009 & 0.0021 \\ \hline
\multicolumn{9}{|l|}{$z$ bin 3}                                                                                                                                                                                                                                                                                                                                                                                                                                                       \\ 
\hline
0.1200 & 0.0034 & 0.0275 & 0.0054 & 0.0259 & 0.0368 & 0.0060 & 0.0603 & 0.0090 \\ \hline
0.2000 & 0.0163 & 0.0192 & 0.0187 & 0.0180 & 0.0556 & 0.0053 & 0.0612 & 0.0047 \\ \hline
0.2800 & 0.0463 & 0.0173 & 0.0477 & 0.0161 & 0.0666 & 0.0046 & 0.0675 & 0.0034 \\ \hline
0.4400 & 0.0417 & 0.0145 & 0.0421 & 0.0134 & 0.0475 & 0.0036 & 0.0781 & 0.0030 \\ \hline
0.6000 & 0.0425 & 0.0172 & 0.0408 & 0.0153 & 0.0271 & 0.0030 & 0.1100 & 0.0043 \\ \hline
0.6800 & 0.0640 & 0.0207 & 0.0589 & 0.0182 & 0.0221 & 0.0032 & 0.1220 & 0.0054 \\ \hline
0.7600 & 0.0632 & 0.0283 & 0.0573 & 0.0240 & 0.0238 & 0.0039 & 0.1512 & 0.0079 \\ \hline
\multicolumn{9}{|l|}{$x-Q^2$ bin 2}                                                                                                                                                                                                                                                                                                                                                                                                                                                       \\ 
\hline
\multicolumn{9}{|l|}{$z$ bin 1}                                                                                                                                                                                                                                                                                                                                                                                                                                                       \\ 
\hline
0.0400 & -0.0105 & 0.0319 & -0.0096 & 0.0293 & 0.0008 & 0.0090 & 0.0822 & 0.0070 \\ \hline
0.1200 & 0.0039 & 0.0164 & 0.0045 & 0.0151 & 0.0106 & 0.0043 & 0.0810 & 0.0034 \\ \hline
0.2000 & 0.0554 & 0.0130 & 0.0518 & 0.0119 & 0.0143 & 0.0031 & 0.0868 & 0.0027 \\ \hline
0.2800 & 0.0586 & 0.0121 & 0.0554 & 0.0110 & 0.0227 & 0.0030 & 0.0890 & 0.0025 \\ \hline
0.3600 & 0.0553 & 0.0115 & 0.0524 & 0.0104 & 0.0259 & 0.0031 & 0.0978 & 0.0024 \\ \hline
0.4400 & 0.0668 & 0.0128 & 0.0632 & 0.0116 & 0.0289 & 0.0031 & 0.0955 & 0.0026 \\ \hline
0.5200 & 0.0424 & 0.0148 & 0.0425 & 0.0134 & 0.0427 & 0.0039 & 0.0929 & 0.0030 \\ \hline
0.6000 & 0.0944 & 0.0188 & 0.0892 & 0.0172 & 0.0328 & 0.0050 & 0.0843 & 0.0037 \\ \hline
0.6800 & 0.0821 & 0.0287 & 0.0789 & 0.0264 & 0.0418 & 0.0078 & 0.0797 & 0.0053 \\ \hline
\multicolumn{9}{|l|}{$z$ bin 2}                                                                                                                                                                                                                                                                                                                                                                                                                                                       \\ 
\hline
0.0400 & -0.0112 & 0.0151 & -0.0105 & 0.0138 & -0.0022 & 0.0050 & 0.0869 & 0.0034 \\ \hline
0.1200 & 0.0164 & 0.0079 & 0.0164 & 0.0072 & 0.0160 & 0.0025 & 0.0825 & 0.0017 \\ \hline
0.2000 & 0.0499 & 0.0066 & 0.0481 & 0.0061 & 0.0276 & 0.0021 & 0.0832 & 0.0012 \\ \hline
0.2800 & 0.0533 & 0.0057 & 0.0516 & 0.0052 & 0.0333 & 0.0020 & 0.0859 & 0.0011 \\ \hline
0.3600 & 0.0632 & 0.0062 & 0.0602 & 0.0056 & 0.0320 & 0.0020 & 0.0947 & 0.0011 \\ \hline
0.4400 & 0.0760 & 0.0068 & 0.0715 & 0.0061 & 0.0322 & 0.0020 & 0.1034 & 0.0012 \\ \hline
0.5200 & 0.0650 & 0.0072 & 0.0620 & 0.0064 & 0.0391 & 0.0024 & 0.1157 & 0.0015 \\ \hline
0.6000 & 0.0785 & 0.0091 & 0.0733 & 0.0080 & 0.0376 & 0.0026 & 0.1253 & 0.0019 \\ \hline
0.6800 & 0.0740 & 0.0108 & 0.0696 & 0.0093 & 0.0418 & 0.0034 & 0.1358 & 0.0027 \\ \hline
0.7600 & 0.0842 & 0.0150 & 0.0786 & 0.0128 & 0.0450 & 0.0043 & 0.1431 & 0.0038 \\ \hline
\multicolumn{9}{|l|}{$z$ bin 3}                                                                                                                                                                                                                                                                                                                                                                                                                                                       \\ 
\hline
0.0400 & 0.0017 & 0.0267 & 0.0035 & 0.0241 & 0.0200 & 0.0077 & 0.0984 & 0.0072 \\ \hline
0.1200 & 0.0173 & 0.0158 & 0.0191 & 0.0144 & 0.0376 & 0.0049 & 0.0891 & 0.0037 \\ \hline
0.2000 & 0.0390 & 0.0128 & 0.0404 & 0.0116 & 0.0529 & 0.0041 & 0.0991 & 0.0029 \\ \hline
0.2800 & 0.0607 & 0.0127 & 0.0617 & 0.0114 & 0.0704 & 0.0041 & 0.1042 & 0.0026 \\ \hline
0.3600 & 0.0492 & 0.0124 & 0.0511 & 0.0109 & 0.0654 & 0.0041 & 0.1205 & 0.0027 \\ \hline
0.4400 & 0.0855 & 0.0139 & 0.0830 & 0.0120 & 0.0672 & 0.0042 & 0.1344 & 0.0031 \\ \hline
0.5200 & 0.0325 & 0.0162 & 0.0377 & 0.0135 & 0.0651 & 0.0043 & 0.1607 & 0.0041 \\ \hline
0.6000 & 0.0835 & 0.0218 & 0.0791 & 0.0178 & 0.0588 & 0.0047 & 0.1804 & 0.0056 \\ \hline
0.6800 & 0.0917 & 0.0325 & 0.0834 & 0.0256 & 0.0521 & 0.0054 & 0.2109 & 0.0087 \\ \hline
\multicolumn{9}{|l|}{$x-Q^2$ bin 3}                                                                                                                                                                                                                                                                                                                                                                                                                                                       \\ 
\hline
\multicolumn{9}{|l|}{$z$ bin 1}                                                                                                                                                                                                                                                                                                                                                                                                                                                       \\ 
\hline
0.1200 & 0.0746 & 0.0162 & 0.0687 & 0.0147 & 0.0105 & 0.0049 & 0.0931 & 0.0043 \\ \hline
0.2000 & 0.0431 & 0.0131 & 0.0410 & 0.0117 & 0.0227 & 0.0036 & 0.1024 & 0.0035 \\ \hline
0.2800 & 0.0474 & 0.0122 & 0.0448 & 0.0108 & 0.0244 & 0.0034 & 0.1121 & 0.0034 \\ \hline
0.3600 & 0.0560 & 0.0122 & 0.0535 & 0.0107 & 0.0349 & 0.0035 & 0.1204 & 0.0035 \\ \hline
0.4400 & 0.0564 & 0.0142 & 0.0542 & 0.0125 & 0.0380 & 0.0040 & 0.1188 & 0.0038 \\ \hline
0.5200 & 0.0697 & 0.0160 & 0.0671 & 0.0141 & 0.0484 & 0.0046 & 0.1223 & 0.0043 \\ \hline
0.6000 & 0.0480 & 0.0214 & 0.0476 & 0.0189 & 0.0451 & 0.0059 & 0.1184 & 0.0053 \\ \hline
0.6800 & 0.0673 & 0.0345 & 0.0645 & 0.0302 & 0.0449 & 0.0087 & 0.1243 & 0.0073 \\ \hline
0.7600 & 0.0867 & 0.0528 & 0.0811 & 0.0476 & 0.0297 & 0.0150 & 0.0978 & 0.0109 \\ \hline
\multicolumn{9}{|l|}{$z$ bin 2}                                                                                                                                                                                                                                                                                                                                                                                                                                                       \\ 
\hline
0.0400 & 0.0219 & 0.0176 & 0.0195 & 0.0153 & 0.0038 & 0.0058 & 0.1296 & 0.0059 \\ \hline
0.1200 & 0.0290 & 0.0098 & 0.0279 & 0.0086 & 0.0199 & 0.0032 & 0.1163 & 0.0031 \\ \hline
0.2000 & 0.0381 & 0.0077 & 0.0371 & 0.0068 & 0.0304 & 0.0027 & 0.1242 & 0.0024 \\ \hline
0.2800 & 0.0510 & 0.0074 & 0.0493 & 0.0065 & 0.0378 & 0.0026 & 0.1277 & 0.0022 \\ \hline
0.3600 & 0.0533 & 0.0077 & 0.0522 & 0.0067 & 0.0453 & 0.0027 & 0.1355 & 0.0022 \\ \hline
0.4400 & 0.0720 & 0.0082 & 0.0690 & 0.0070 & 0.0512 & 0.0029 & 0.1474 & 0.0025 \\ \hline
0.5200 & 0.0729 & 0.0101 & 0.0694 & 0.0085 & 0.0515 & 0.0032 & 0.1631 & 0.0030 \\ \hline
0.6000 & 0.0897 & 0.0132 & 0.0832 & 0.0109 & 0.0521 & 0.0037 & 0.1726 & 0.0037 \\ \hline
0.6800 & 0.0952 & 0.0153 & 0.0894 & 0.0126 & 0.0623 & 0.0045 & 0.1772 & 0.0048 \\ \hline
0.7600 & 0.0995 & 0.0233 & 0.0915 & 0.0191 & 0.0553 & 0.0051 & 0.1789 & 0.0067 \\ \hline
\multicolumn{9}{|l|}{$z$ bin 3}                                                                                                                                                                                                                                                                                                                                                                                                                                                       \\ 
\hline
0.1200 & -0.0469 & 0.0402 & -0.0381 & 0.0353 & 0.0245 & 0.0059 & 0.1237 & 0.0120 \\ \hline
0.2000 & 0.0728 & 0.0324 & 0.0693 & 0.0275 & 0.0496 & 0.0050 & 0.1503 & 0.0100 \\ \hline
0.2800 & 0.0716 & 0.0323 & 0.0707 & 0.0270 & 0.0657 & 0.0050 & 0.1611 & 0.0097 \\ \hline
0.3600 & 0.0542 & 0.0360 & 0.0563 & 0.0295 & 0.0656 & 0.0049 & 0.1810 & 0.0108 \\ \hline
0.4400 & 0.1026 & 0.0427 & 0.0967 & 0.0343 & 0.0726 & 0.0050 & 0.1966 & 0.0126 \\ \hline
0.5200 & 0.1249 & 0.0564 & 0.1127 & 0.0442 & 0.0686 & 0.0054 & 0.2165 & 0.0167 \\ \hline
0.6000 & 0.0281 & 0.0833 & 0.0394 & 0.0595 & 0.0676 & 0.0057 & 0.2859 & 0.0265 \\ \hline
\caption{Result of the 4-dimensional study with $z$ or $P_T$ as the main bin variable with 3 $x_B-Q^2$ bins and 3 $P_T$ or $z$ bins. The column $z$/$P_T$ provides the bin middle. The other columns provide the value and the statistical uncertainty of the corrected kaon, the raw kaon and the pion $F_{LU}^{\sin{\phi}}/F_{UU}$ and the pion contamination in the kaon sample estimated from MC.}
\end{longtable}

\bibliography{Kaon_paper_bib}